\documentclass[twocolumn]{aastex631}

\shorttitle{Inflated Eccentric Migration of Evolving Gas-Giants I}
\shortauthors{Rozner et al}

\usepackage{graphicx}	
\usepackage{amsmath}	
\usepackage{amssymb}	
\usepackage{float}
\usepackage{ulem}
\usepackage{soul}
\usepackage{color}
\definecolor{Blue}{rgb}{0.25,.41,.88}
\definecolor{Red}{rgb}{0.7,.3,.0}

\begin{document}


\title{Inflated Eccentric Migration of Evolving Gas-Giants I: \\Accelerated Formation and Destruction of Hot and Warm Jupiters}


\email{morozner@campus.technion.ac.il}

\author[0000-0002-2728-0132]{Mor Rozner}
\affiliation{Technion - Israel Institute of Technology, Haifa, 3200002, Israel}

\author[0000-0002-6012-2136]{Hila Glanz}
\affiliation{Technion - Israel Institute of Technology, Haifa, 3200002, Israel}


\author[0000-0002-5004-199X]{Hagai B. Perets}
\affiliation{Technion - Israel Institute of Technology, Haifa, 3200002, Israel}

\author[0000-0001-7113-723X]{Evgeni Grishin}
\affiliation{Technion - Israel Institute of Technology, Haifa, 3200002, Israel}
\affiliation{School of Physics and Astronomy, Monash University, Clayton 3800, VIC, Australia}

\begin{abstract}
Hot and warm Jupiters (HJs  and  WJs  correspondingly) are gas-giants orbiting their host stars at very short orbital periods ($P_{HJ}<10$ days; $10<P_{WJ}<200$ days). HJs and a significant fraction of WJs are thought to have migrated from an initially farther-out birth locations. While such migration processes have been extensively studied, the thermal evolution of gas-giants and its coupling to the the migration processes are usually overlooked. In particular, gas-giants end their core-accretion phase with large radii and then contract slowly to their final radii. Moreover, intensive heating can slow the contraction at various evolutionary stages. The initial inflated large radii lead to faster tidal migration due to the strong dependence of tides on the radius.  Here we explore this accelerated migration channel, which we term inflated eccentric migration, using a semi-analytical self-consistent modeling of the thermal-dynamical evolution of the migrating gas-giants, later validated by our numerical model (see a companion paper, paper II). We demonstrate our model for specific examples and carry a population synthesis study. Our results provide a general picture of the properties of the formed HJs\&WJs via inflated migration, and the dependence on the initial parameters/distributions. We show that tidal migration of gas-giants could occur far more rapidly then previously thought and lead to accelerated destruction and formation of HJs and enhanced formation rate of WJs. Accounting for the coupled thermal-dynamical evolution is therefore critical to the understanding of HJs/WJs formation, evolution and final properties of the population and play a key role in their migration process. 
\end{abstract}



\section{Introduction}
Formation of gas-giant planets is thought to occur either through a gradual bottom-up growth of planetary embryos followed by rapid gas accretion (core accretion) or through a direct formation of gas-giants following gravitational-instability \citep{Mizuno1980,Boss1997,Armitage2010}. However, the observed proximity of hot and warm Jupiters (HJs and WJs correspondingly) to their host star sets severe constraints on both these mechanisms. The high temperatures, high velocities, low disc-mass and solids that characterize these environments constrict the predicted formation rates from the in-situ channel \citep{Bodenheimer2000,Rafikov2005}. Though in-situ formation might still potentially explain a non-negligible fraction of the population of {\rm warm} Jupiters residing farther-out \citep{HuangWuTriad2016,AndersonLaiPu2020}, formation of HJs and eccentric WJs is not likely to occur in-situ. These were therefore suggested to have formed at larger separations from the star, and then migrated inwards to their current locations (see \citealp{Dawson2018} for a review). 

Currently suggested migration channels can be generally divided between disc-migration channels and high eccentricity tidal-migration channels \citep{Dawson2018}. While such migration scenarios could potentially explain the origins of HJs\&WJs, detailed studies of these migration scenarios struggled to reproduce the rates and properties of formed HJs\&WJs \citep{Dawson2018,ZhuDong2021}. 
However, such models usually did not self-consistently account for role of the thermal evolution of the gas-giants, their initially hot inflated state and later contraction (and possible heating) during their eccentric-migration.  

In high eccentricity migration mechanisms, the semimajor axis shrinks due to energy dissipation induced by tidal forces.  This process requires initial high eccentricity, as tidal coupling strongly depends on the distance from the star, which becomes small at pericenter approaches of highly eccentric orbits. The planet can initially be excited to such high eccentricities either by planet-planet scattering \citep{RasioFord1996,Chatterjee2008,JuricTremaine2008}, through Von-Ziepel - Lidov - Kozai mechanism \citep{VonZeipel1910,Lidov1962,Kozai1962} coupled to tidal-evolution \citep[e.g.][]{KiselevaEggletonMikkola1998,WuMurray2003,FabryckyTremaine2007,Naoz2011HJ,Petrovich2015a,gpf18,VickLaiAnderson2019} or coupled multiplanet secular evolution \citep[e.g.][]{WuLithwick2011,Beauge2012,HamersLithwichPeretsZwart2017}
and their combinations. In this paper we focus on excitation by planet-planet scattering, and we will discuss the secular channel in a future study. Current tidal migration models show that typical migration timescales could be long, and the production rate HJs\&WJs (or both) are too low in comparison with observations (and in particular for WJs). However, as we discuss below, the initial hot-inflated state of gas-giant at birth and their later radiative and tidal heating could change this picture.  

Core accretion formation of gas-giants proceeds through the runaway accretion of gas from the protoplanetary disk onto a typically few Earth mass solid core \citep{Perri1974,BodenheimerPollack1986}. At the end of core accretion, the recently formed gas-giants reach large radii possibly up to $10 \  R_J$, for dusty planets \citep{GinzburgChiang2019}. After a very rapid contraction phase, when the planet's radius contracts to $~4 \ R_J$ \citep{Guillot1996}, the contraction slows down, and takes place at typical Kelvin-Helmholtz timescales (tens of $\rm{Myrs}$) as planets contract asymptotically to reach their (effective) final size. The contraction depends on the mass and, as we discuss in depth, on the externally injected energies due to tides and/or irradiation, with only weak dependence on the exact initial conditions at birth (leaving uncertainties/degeneracy in the models of early evolution of gas-giants, e.g. \citealp{Marley2007}). It should be noted that dustier planets could lead to even slower contraction \citep{GinzburgChiang2019}. 

Typically, eccentric migration scenarios assume the radii of the migrating gas-giants are constant throughout their evolution, and is taken to be their asymptotic radii at late times, $\sim 1 R_J$, neglecting the initial inflated radii at birth and later contraction.  Since tidal forces strongly depend on the radius of the affected object, initially inflated planets should migrate faster than planets which already contracted to their final radii. The dynamical and thermal evolution can not be decoupled, due to their mutual strong dependence on each other.  To date, these issues were only partially studied by \cite{Wu2006,MillerFortneyJackson2009,Petrovich2015a}, and even these only addressed limited aspects of coupled evolution models using simplified approaches. 

In this paper and Glanz, Rozner, Perets $\&$ Grishin 2021 \citealt{Glanz2021} (hereafter paper II) we study the coupled dynamical-thermal evolution of HJs\&WJs, which evolve via inflated eccentric migration, after the end of the core accretion phase, and compare them with corresponding cases of initially non-inflated planets, and inefficient heat conduction typically studied in the literature. We present a semi-analytical model, described in this paper and compare it with a numerical model, discussed in paper II. Both approaches couple the dynamical and tidal evolution of the planets with their thermal evolution. The former makes use of a simple analytic approach for the thermal evolution and its effect on the radius evolution, and the latter follows the evolution through a numerical model using the stellar and planetary evolution code \texttt{MESA} \citep{PaxtonMESA2011,PaxtonMESA2013}, coupled with the \texttt{AMUSE} framework \citep{Portegies2009AMUSE}. We find an excellent agreement between the models results in terms of the overall evolution and properties of the systems, as we discuss below and in paper II. Consequently, and due to the simplicity of the semi-analytical approach relative to the computationally costly numerical one, we discuss results from the numerical evolution only for specific cases, and show comparisons to the analytic model (see below and in paper II). 
The efficient semi-analytic approach allows us to study the large parameter-space of possible initial conditions and study the formed population of HJs/WJs, which are inaccessible to the numerical approach, which is computationally expensive and has limitations in long timescales/small radii (see Paper II for further details).

In the following we describe our semi-analytic approach, and present the various aspects of inflated eccentric-migration and its outcomes. We then consider the overall distribution of initial conditions and the resulted rates and properties of formed HJs\&WJs and their dependence on these initial distributions.

In order to present our approach we first discuss the conditions for inflated eccentric migration arising from planet-planet scattering. We then discuss tidal migration in general, and elaborate about weak and dynamical tidal models \ref{sec:high eccentricity tidal migration} and their use in our models. In section \ref{sec:semi-analytical} we describe the semi-analytical model for inflated migration and demonstrate the formation of HJs and WJs in specific examples. 
In section \ref{sec:case study} we demonstrate the use of the semi-analytical model for several examples and compare to the numerical model. 
In section \ref{sec:Monte-Carlo} we discuss population synthesis of HJs and WJs formation. 
In section \ref{sec:reults}, we present the results of the population synthesis, the choice of its parameters and the role played by them. 
In section \ref{sec:discussion & implications} we discuss our findings and further implications. In section \ref{sec:summary} we summarize our results. 

\section{Planet-planet scattering}
In this study we focus on planetary systems in which the initial conditions for high eccentricity migration are dictated by planet-planet scattering.  
There are other channels for eccentricity excitation,
such as secular Lidov-Kozai evolution in triple systems   \citep[e.g.][]{WuMurray2003,Petrovich2015a,Petrovich2015b,Naoz2011HJ,gpf18,VickLai2018}, and secular resonances \citep[e.g][]{WuLithwick2011,HamersLithwichPeretsZwart2017}. Such processes typically operate on longer timescales and lead to intermittent/quasi-periodic high eccentricities and are also sensitive to precession induced by tidal interactions which can partially quench the level of eccentricity excitation. In this work we consider only planet-planet scattering, but inflated eccentric migration is likely to be important for those other channels such as the ones involves secular evolution \citep[see ][]{Wu2006,Petrovich2015a}. It should be noted that in general the timescales of these channels could be longer, suggesting that a priori the effect of initial inflation might be smaller. However, initially inflated planets will still leave a signature on the dynamical evolution. First, a larger fraction of the initial planets are prone to tidal disruption. Moreover, the final distribution of the semimajor axes is expected to change accordingly \citep{Petrovich2015a}.
Another potential effect of initial inflation is a tidal quenching of the LK mechanism, that passes with the contraction of the planet.
Although several studies were done on the secular channel, to our knowledge the coupling there between the thermal and dynamical is not self-consistent; however, a self-consistent modeling could be implied easily in our semi-analytical model.
This channel is out of the scope of this paper and is left out for followup studies.

In multi-planet systems, mutual gravitational interactions between the planets perturb their orbit and may destabilize the system leading to ejection of planets, mutual collisions, collisions with the star and more generally excitation of the planets eccentricities and inclinations \citep{RasioFord1996,WeidenschillingMarzari1996,Chatterjee2008}. The gravitational encounters could also involve tidal circularization during the scattering process, that might enhance the fractions of shorter-period and high eccentricity gas-giants \citep{Nagasawa2008}.
More generally, it was found that strong planet-planet scatterings eventually give rise  to a Rayleigh distribution of the eccentricities of the surviving planets \cite[e.g.][]{JuricTremaine2008}. It should be noted that although good fits exist, the exact final eccentricity distribution is still unknown in analytic terms and usually a preliminary N-body simulation is needed to determine the distribution at the end of planet-planet scattering.  
In our calculations we assume that destabilized systems evolve through planet-planet scattering, leading to such eccentricity distribution, given by
\begin{align}
\frac{dN}{de}\propto e \exp\left[-\frac{1}{2}\left(\frac{e}{2\sigma_e^2}\right)^2\right]
\end{align}
where we adopt a value of $\sigma_e=0.5$ in our fiducial model (see table \ref{table:rates_with_star_formation} and section \ref{sec:Monte-Carlo}) 
We note that some studies considered somewhat different distributions with even higher fractions of highly eccentric orbits, that might improve our results, e.g. \cite{Nagasawa2008,Carrera2019}.

The Rayleigh distribution of the eccentricities arising from planet-planet scattering determines the fractions of planets that would evolve through inflated eccentric migration to become HJ/WJ and would not be tidally disrupted by the host star. 

The tidal disruption radius is given by $r_{\rm dis}=\eta R_{p}\left(M_\star/M_p\right)^{1/3}$ where $\eta=2.7$ \citep{Guillochon2011}, and planets with pericenter approaching these value are assumed to be disrupted and, naturally  are not considered to be HJ/WJs. Planets with too high pericenter approach that are not affected by tides, or those that do migrate through inflated eccentric migration, but do not become HJs or WJs in the relevant time considered, i.e. the age considered for the HJ/WJ, are similarly not considered to be HJs\&WJs. 

The planet-planet scattering timescale sets the initial radius of the planet in the migration stage.
Gas giant usually form with a typical radius that can exceed even $4 \ R_J$, from which there will be a rapid cooling phase up to $4 \ R_J$ \citep{Guillot1996}. The following contraction phase is slower, along the Hayashi track. This timescale could take less than Myr to a typical radius of $1.5 \ R_J$. The decoupling timescale from the planet-planet scattering could stray from less then Myr to few Myrs and even more (e.g. \citealp{Dawson2018} Fig. 3), such that the initial radius for the migration phase could be large (corresponds to short decoupling time) or small (long decoupling time), depending on the planet-planet scattering conditions. A more detailed calculation, setting more accurately the initial distribution for our semi-analytical based population synthesis, is out of the scope of this paper and is left out for a future study.
However, we do take this into account by considering several possible initial radii distribution.

\section{High Eccentricity Tidal Migration}\label{sec:high eccentricity tidal migration}

Tidal migration is a dissipative process, where tides raised on the planet by the host star extract energy from the planet orbit, typically leading to its inward migration into shorter period orbits. Tides raised on the star by the planet can also contribute to tidal migration, but these are typically negligible compared to the effects of tides raised on the planet, though they might become important under some circumstances \citep{GinzburgSari2017}. In the following we only consider tides raised on the planet and postpone the consideration of tides on the star to later studies. These will lead to corrections in eq. \ref{eq:weak tide} may be included
(e.g. \citealp{MillerFortneyJackson2009}).

In order for the tidal dissipation to be effective and lead to a significant migration, a small pericenter approach should be considered. 
This dictates the basic initial conditions for eccentric migration. High eccentricity tidal migration could therefore be divided roughly into two separate steps: reducing the planet's angular momentum and reducing the planet's energy. In the first step, the HJ/WJ progenitor is excited into a an eccentric orbit via planet-planet scattering \citep{RasioFord1996,Chatterjee2008,JuricTremaine2008}, as we discuss here, or through other channels for eccentricity excitation (e.g. via Von-Zeipel - Lidov - Kozai mechanism \citealp{VonZeipel1910,Lidov1962,Kozai1962,WuMurray2003,FabryckyTremaine2007,Naoz2011HJ,Petrovich2015a}).  In the second step, 
energy extraction via tides leads to a migration and circularization of the planet's orbit. 
The energy extracted from the orbit (calculated over an orbital period) per period is dissipated in the planet, and therefore the injected energy heat the planet. The injected energy per unit time,
is given by, 
\begin{align}\label{eq:L_tide}
L_{\rm tide} = -\frac{E}{a}\frac{da}{dt}
\end{align}

where $E$ is the orbital energy and $a$ is the semimajor axis. 
The angular momentum is approximately conserved. For HJs the final orbit is usually circular, and given angular momentum conservation one can estimate the final semimajor axis of the HJ, finding $a_{\rm final}=a_{0}(1-e^2_0)$, where $a_0$ and $e_0$ are the initial semimajor axis and eccentricity correspondingly. The same consideration could be applied for any other given final eccentricity (or any eccentricity during the evolution).

Modeling tides is not trivial; in particular,  their strong dependence on the internal structure of the planet and other physical aspects of the problem raise many complications. Here we consider two tide models, one is the widely used tidal model of weak/equilibrium tides \citep{Darwin1879,GoldreichSoter1966,Alexander1973,Hut1981}, the second one is a model for dynamical tides \citep{Zahn1977, Mardling1995a,Mardling1995b}. The latter could be especially important and more efficient during the early migration phases when the planet orbit is still highly eccentric, and in that sense, considering only weak-tides model is potentially conservative in term of the efficiency of eccentric migration and long resulted timescales \citep{Lai1997}.
Our approach is general and any other tide model could be incorporated potentially such as the chaotic-dynamical tides \citep{IvanonPapaloizou2004,IvanovaPapaloizou2007,VickLai2018,Wu2018,VickLaiAnderson2019}, which potentially further shorten the migration timescales due to more efficient extraction of energy. The tide models are discussed below.  Here we consider only non-chaotic dynamical tides.
It should be noted that the very same equations are used in the detailed numerical model in paper II. To keep this paper is succinct as possible and to minimize overlap between it and paper II, we refer the reader to paper II for further discussion on the tidal evolution, which is relevant also to the semi-analytical model.

\subsection{Equilibrium tide model}\label{subsec: equilibrium tide model}

The equilibrium tide model is used in many physical scenarios and is widely discussed (e.g.
\cite{Darwin1879,GoldreichSoter1966,Alexander1973,Hut1981}). 

Let us consider an orbit-averaged time evolution of the eccentricity and semimajor axis. If we assume that pseudo-synchronization of the planetary spin and the orbit occurs on short time scale, and that the angular momentum is conserved during the migration, one finds that \citep[e.g.][]{Hut1981,HamersTremaine2017}

\begin{align}\label{eq:weak tide}
\frac{da}{dt}=& -21 k_{\rm AM} n^2 \tau_p \frac{M_\star}{M_p}\left(\frac{R_p}{a}\right)^5ae^2 \frac{f(e)}{(1-e^2)^{15/2}}, \\
\frac{de}{dt} =& -\frac{21}{2}k_{\rm AM}n^2 \tau_p \frac{M_\star}{M_p} \left(\frac{R_p}{a}\right)^5 e\frac{f(e)}{(1-e^2)^{13/2}}
\end{align}
where $M_\star$ is the mass of the host star, $M_p, \ R_p, \ e, \ a$ and $n$ are the mass, radius, orbital eccentricity, orbital semimajor axis and mean motion of the gas-giant correspondingly, $\tau_p=0.66 \  \rm sec$ is the planetary tidal lag time and $k_{\rm AM}=0.25$ is the planetary apsidal motion constant ($k_{\rm AM}$ and $\tau_p$ are taken from \citealp{HamersTremaine2017}). 
$f(e)$ is defined by 

\begin{align}
f(e) = \frac{1+\frac{45}{14}e^2+8e^4+\frac{685}{224}e^6+\frac{255}{448}e^8+\frac{25}{1792}e^{10}}{1+3e^2+\frac{3}{8}e^4}
\end{align}

The energy extracted per period and hence the migration and circularization rate, scales as $R_p^5$. Consequently, the migration timescales of initially inflated gas-giants, where the planetary radius, $R_p>R_J$, are shortened relative to migration of non-inflated gas-giants, i.e. with a constant radius of $R_J$. The contraction timescales are long enough to maintain inflated gas-giants along a significant part of their dynamical evolution, such that the initial radius of a HJ/WJ will leave a signature on its expected final parameters, that could be also observed.

\subsection{Dynamical Tides}\label{subsec: dynamical tides}
Tidal forcing from the star might excite internal energy modes of the planet (mainly the fundamental f-mode), which might induce an enhanced response \citep{Mardling1995a,Mardling1995b,Lai1997,Ogilvie2014}. The energy is mostly extracted during  the pericenter approach and the extraction is more efficient compared with the equilibrium tide model, potentially leading to even more rapid circularization and migration of the planet. The eccentricity decay is accompanied by pseudo-synchronization with the angular frequency of the star, and the excitation of oscillations in the planet become less pronounced as the orbital eccentricity decrease. The energy dissipation by the various modes is gradually suppressed, until a transition to the regime in which equilibrium tides become more dominant. The quadrupole order of the energy dissipation can be written as follows \citep{PressTeukolsky1977,MoeKratter2018}, 

\begin{align}\label{eq: dynamical tides}
\Delta E = f_{\rm dyn}\frac{M_\star+M_p}{M_p}\frac{GM_\star^2}{R_p}\left(\frac{a(1-e)}{R_p}\right)^{-9}
\end{align}
with $f_{\rm dyn}=0.1$ unless stated otherwise \citep{MoeKratter2018} 
taken for our case of tidal response of a gas-giant. 
Combining this prescription with the equations of the orbital energy and angular momentum, and assuming a constant pericenter, leads to the following equations of the orbital semimajor axis and eccentricity along the migration \citep{MoeKratter2018}, 

\begin{align}\label{eq:dynamical tides}
\frac{da}{dt} = \frac{a}{P} \frac{\Delta E}{E}, \ 
\frac{de}{dt}= \frac{1-e}{a}\frac{da}{dt}
\end{align}

While dynamical tides dominate for large eccentricities, weak tides will be a more physical description for low ones.
The ratio of the migration rate due to dynamical tides to the migration rate due to weak tides is given by 
\begin{align}
\beta(R_p,a,e) :=\frac{da/dt|_{\rm dyn}}{da/dt|_{\rm weak}}= \frac{2f_{\rm dyn}R_p^3(1-e^2)^{15/2}}{21 GM_p k_{\rm AM}\tau_p (1-e)^9  Pe^2 f(e)}
\end{align}
where $P$ is the period of the planet. 
The transition between the dynamical and weak tides occurs roughly at $\beta \sim 1$, and we set a lower artificial cutoff at $e=0.2$, such that the transition occurs at $\max\{0.2,e|_{\beta=1}\}$, to avoid the divergence of dynamical tides at $e=0$.

\section{Combined Thermal-Dynamical Semi-Analytical Model}\label{sec:semi-analytical}
We model inflated eccentric migration by coupling the orbital equations, governed by the tidal migration model, with the thermal evolution of the planet, dictated by heating due to tides and irradiation and thermal cooling. The thermal evolution leads to a planetary contraction due to cooling, which can be slowed by external heating sources and in extreme cases -- even be stopped/partially reversed. The tidal evolution is strongly affected by the radius of the planet, such that the thermal evolution changes the migration rate and timescale.

The virial theorem states a relation between the total and the potential energies $E=-(3\gamma-4)U/(3\gamma-3)$, such that $U\propto 3GM_p^2/R(5-\tilde n)$ where $\tilde n=1/(\gamma-1)$ is the polytropic index,
$\gamma$ is the heat capacity ratio, taken as $\gamma=5/3$ and the proportion constant is determined by numerical gauge. The change in the planet luminosity is mostly determined by the change in the thermal energy of ions, which are not degenerate, and their equation of state is given by the ideal gas equation $E=(M_p/\mu)k_BT_c$ where $\mu$ is the molecular weight, taken as the proton mass and $T_c$ is central temperature of the planet. Following this relation, we derive the following equation for the radius change due to heating and cooling, 

\begin{align}\label{eq:dRdt}
\frac{dR_p}{dt} \propto \frac{5\gamma-6}{3\gamma-4}\frac{R_p^2}{GM_p^2}
\left(L_{\rm ext}-L_{\rm cool}\right)\end{align}

It should be noted that the energy deposition and loss are generally not constant and could vary with time and the change of the orbital parameters -- the semi-analytical approach takes that into account. In a similar manner one can calculate the evolution of the central temperature of the planet:
\begin{align}\label{eq:dTcdt}
\frac{dT_c}{dt}= \frac{m_p}{M_p k_B} \left(L_{\rm ext}-L_{\rm cool}\right)
\end{align}
where $m_p$ is the proton mass, $L_{\rm cool}$  describes the cooling of the planet and $L_{\rm extra}$ the further external sources of energies injected, such as tidal heating and irradiation, or any other general source of heating.  
 The external luminosity could be generally a function of the optical depth $\tau_{\rm dep}$ in which it is deposited, or equivalently, of the pressure $P_{\rm dep}$. 
The effect of the deposition is the strongest for $\tau_{\rm dep}=\tau_c$, i.e. at the center, and decreases as $\tau_{\rm dep}$ decreases.
Although deposition at layers has different physical consequences than deposition at the center \citep{SpiegelBurrows2013,YoudinMitchell2010,KomacekYoudin2017}, to our purposes, the energy injected at outer layers could be translated to a reduced deposition at the center, as shown by \citealp{GinzburgSari2016} for the case of powerlaw distrbution.

To re-inflate the gas giant, it could be seen directly from eq. \ref{eq:dRdt}, that the minimal external luminosity added should be comparable to the radiating luminosity. For a tidal disruption, a larger energy is required, of the order of magnitude of the binding energy of the planet.

The reduction can be considered by an overall multiplication factor, depending on the depth of the deposition \citep{GinzburgSari2015,GinzburgSari2016}, and the effective temperature could be inferred by the effect from the irradiation of the host star. The exact multiplication factor depends on the yet not well understood heat transfer processes in the planet (which we discuss in more details below).
The planet cools through a blackbody emission, which is given by 
\begin{align}\label{eq:Lcool}
L_{\rm cool}\propto 4\pi R_p^2 \sigma_{\rm SB} T_{\rm eff}^4
\end{align}
  where a correction for blackbody radiation arises from the convective-radiative boundary and the transition from isolation to insolation \citep{GinzburgSari2015}. 
 While deposition at the photosphere doesn't change significantly the cooling rate of the planet, central deposition could lead to more prominent effects. 
 Weak external energy source, such as deposition at the photosphere, could be thought as the 'standard' Kelvin-Helmholtz cooling, up to a small correction rising from the external source, that becomes important when $L_{\rm dep}/L_{\rm cool}\gtrsim 1$. The effective temperature at late times is determined solely by the stellar irradiation and reaches a steady state, with correspondence with the typical observationally inferred values for effective temperatures of HJs and WJs. At early times, the temperature of the gas-giant could exceed this final temperature, but then cool to a quasi-steady temperature and then its temperature rises again to the one expected from stellar irradiation (see for example Figs.\ref{fig:HJ}a, \ref{fig:WJ} a. 
 
While the thermal evolution, and hence also the evolution of the planetary radius, are governed by the central temperature of the planet, the effective temperature, which also appears in the cooling equation, could be changed significantly without changing the central temperature.

The relation between the effective and central temperatures depends on the pressure gradient inside the planet, 
  \begin{align}\label{eq:Teff}
\frac{T_{\rm eff}}{T_c}\propto \left(\frac{P_{\rm RCB}}{P_c}\right)^{\gamma/(\gamma-1)}; \\
P_c \sim \frac{GM_p^2}{R_p^4},\
P_{\rm RCB}\sim \frac{GM_p}{\bar \kappa R_p^2 }
\end{align}
where $P_{c}$ is the central pressure of the planet, $P_{\rm RCB}$ is the pressure at the boundary layer between the radiative and convective regions and $\bar \kappa$ is the mean opacity, averaged over the planet atmosphere. We consider $\bar \kappa =5\times 10^{-2} \ \rm{cm^2/g}$ following the typical values in the literature \citep{GinzburgChiang2019}. The proportion factors could be larger than order of unity, and could be found more precisely from numerical simulations. The photosphere luminosity is determined by the effective boundary condition for $L_{\rm cool}$. In general, the RCB layer is dictated by the internal adiabat of pressures. It is not located at the photosphere, and the photosphere is in fact an underestimation of it (see further discussion in \citealp{Thorngren2019,Sarkis2021}).
It should be noted that  the pressure of the RCB is dictated by both the internal adiabat and the external irradiation, as shown in eq.\ref{eq:Teff}.

The opacity might vary strongly with the temperature change. For high temperature planets, i.e. $T \gtrsim 10^4 \ \rm{K}$, the opacity is given by Kramers bound-free opacity law $\kappa \propto \rho T^{-3.5}$ ($H^-$ opacity) \citep{Kippenhahn2012,GinzburgSari2015}. Moreover, dustier planets could have higher initial opacity, and hence finish their core accretion with larger radii, that remain inflated for longer timescales, enhancing the effect discussed in this paper. 

Coupling equations \ref{eq:dRdt},\ref{eq:dTcdt} with orbit-averaged evolution equations defined by the tidal migration model (i.e. for weak tides is given by eq. \ref{eq:weak tide} and for dynamical tides by eq. \ref{eq:dynamical tides}), provides a complete consistent semi-analytical description of the coupled dynamical-thermal model. 

In the following we discuss the addition of external heating sources, which are discussed in more detail in paper II.

\subsection{External Heating}

Our semi-analytical model is general and could effectively take account of any coupling between the thermal and dynamical evolution, and include external heating sources such as tidal heating and irradiation. The efficiency of heat deposition due to such sources depends both on the heating rate and on the depth of the deposition.  Heat deposition in the central parts of a planet leads to more significant effect compared with a far limited effect of  deposition at the photosphere, where radiative cooling effectively dispose much of the heating.

External heating plays a dual role, maintaining the planet inflated for longer time (/in extreme cases -- inflation) and modifying the effective temperature. At late stages, the observed effective temperature of a gas-giant is determined by the the irradiation/other external energy sources applied on the planet, such that external heating, even if not playing a role in inflation, should be taken into consideration in the effective temperature calculation. 
The relative role of external heating can change during the planet thermal-dynamical evolution. When the planet is sufficiently far away from the star, the heating due to  irradiation is smaller than the internal heat and hence it negligibly affects the thermal evolution. 
When the planet migrates closer to the star, the radiative heating (and tidal heating when applicable) becomes larger and then one needs to consider irradiation when determining the planet temperature. This can be determined by the planetary equilibrium temperature: 
\begin{align}
T_{\rm eff,irr}(t)\propto \left(\frac{L_\star}{16\pi \sigma_{\rm SB}r^2(t) }\right)^{1/4}
\end{align}

\noindent
where $r(t)$ is the instantaneous distance of the planet from the star. In order to include it self-consistently in an efficient way in our model, we consider the averaged effective temperature over an orbit (given that the heating rate is sufficiently small during dynamical times). 
Similar considerations would apply to the effect of other heating sources on the effective temperature. 

\section{Case Study Examples}\label{sec:case study}

In the following we present results of the semi-analytical model and compare between the results from the numerical model, while we refer the reader to paper II for further examples of this comparison, an extensive description of the numerical approach, its limitations and applications.   

 The semi-analytical model could potentially take into account any general distribution of external energies, by integration of contributions from different optical depths (see \citealp{GinzburgSari2016} for the case of a constant energy source with a power-law distribution in the optical depth). 
Currently, for simplicity, the injection of heat is simply put as a point source in the center, while in the numerical modeling one needs to distribute the energy injection over some finite region, as not lead to unstable or divergent result from the nonphysical injection at a singular point. The current chosen distribution in the numerical prescription is then a semi-Gaussian (with dispersion changing with the location of the deposition), that injects heat with a deviation of $15-50\%$ from the intended value (see paper II for the technical details regarding the distribution). 
  It should be noted that the actual physical distribution is not yet understood, since the external heating sources are likely to lead to a distribution with a preference towards the outer areas of the planet, and its extension depends on the exact nature of heat transfer. Nevertheless, since the effect of a generally distributed external energy on the dynamical evolution could be translated to some appropriate amount of energy injected at the center, we practically able to treat the effect of any external energy distribution, through an effective heat injection in the center, with some order-unity constant pre-factor to calibrate between the distributed injection in the numerical model and the semi-analytic one. 

\begin{figure*}
\centering
\includegraphics[width=\linewidth]{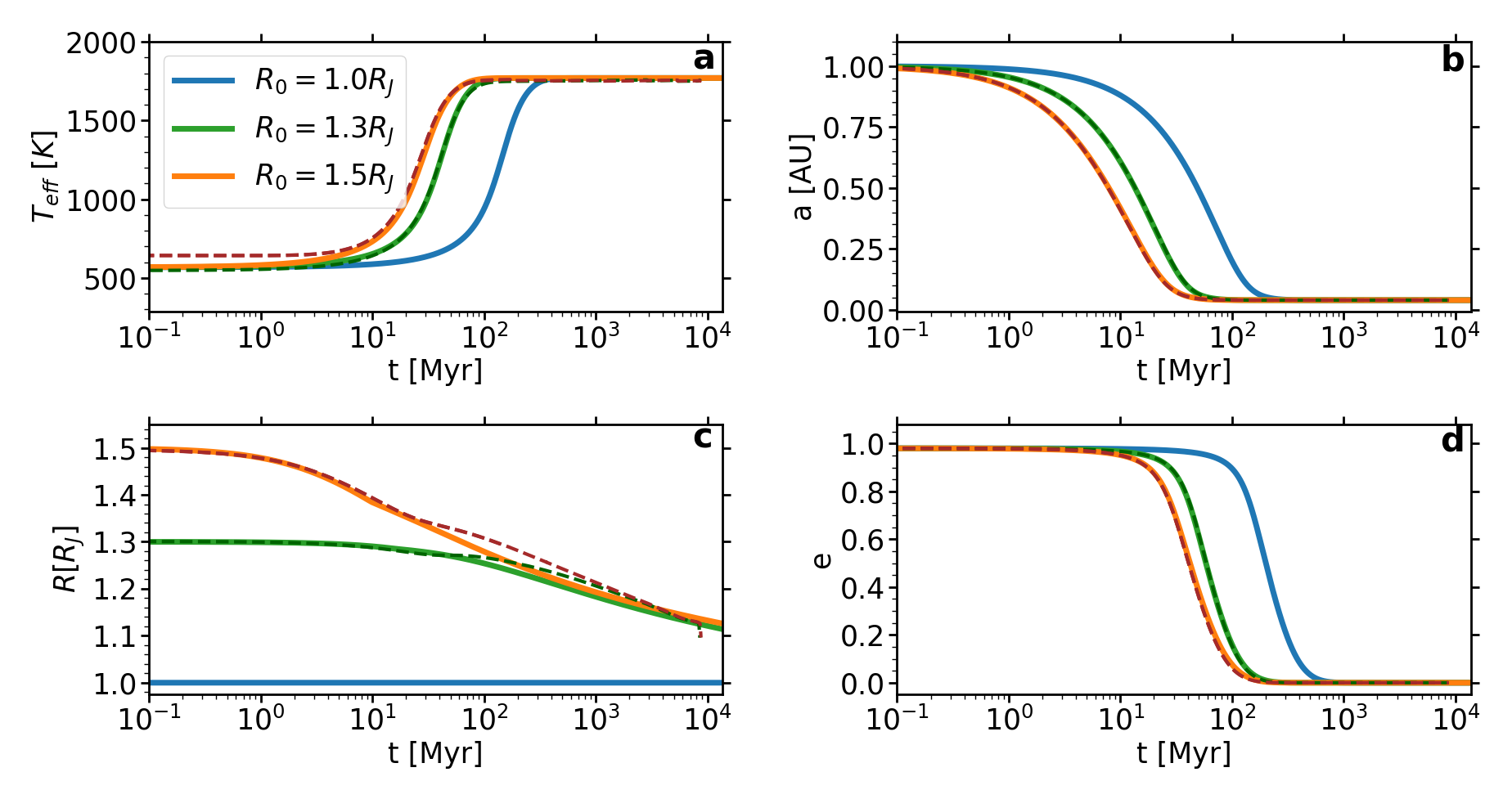}
\caption{The thermal and orbital evolution of a HJ-progenitor migrating due to weak tides, with photosphere heating due to irradiation.  
The thermal (\textbf{a,c}- effective temperature and radius) and orbital (\textbf{b,d} - semimajor axis and eccentricity) evolution of $1 \  M_J$ HJ progenitors with different initial radii ($1 \ R_J$ -- blue, $1.3 \ R_J$ -- orange and $1.5 \ R_J$ -- green). Tides are modeled through a weak-tide model,  irradiation of the planetary outer layer is included, but with no additional efficient heat conduction to the core. Shown are gas-giants with an initial semimajor axis of $1 \ \rm{AU}$ and initial eccentricity of $0.98$. The solid lines correspond to the semi-analytical calculation and the dashed to the numerical.}
\label{fig:HJ}
\end{figure*}

In Fig. \ref{fig:HJ} we present an example of the coupled dynamical-thermal evolution of eccentric $1 \ M_J$ gas-giants with different initial radii, initial semimajor axis of $1 \ \rm{AU}$ and initial eccentricity of $0.98$. We compare the evolution of a constant $1 \ R_J$, with that of initially inflated planets. All the planets in our model experience photospheric heating induced by stellar irradiation.  

We find the semi-analytic and numerical approaches are in excellent agreement and yield similar results (see additional examples in paper II).

The typical migration and circularization timescales are given correspondingly by (for the weak equilibrium tide model)
\small
\begin{align}
&\tau_{\rm mig,weak}=\frac{a}{|\dot a_{\rm weak}|}= \frac{(1-e^2)^{15/2}M_p}{21 k_{\rm AM} n^2 \tau_p M_\star e^2 f(e)}\left(\frac{a}{R_p}\right)^5\\
&\tau_{\rm circ,weak}= \frac{e}{|\dot e_{\rm weak}|}=\frac{2}{21}\frac{M_p(1-e^2)^{13/2}}{k_{\rm AM}M_\star n^2 \tau_p f(e)}\left(\frac{a}{R_p}\right)^5
\end{align}
\normalsize
\noindent
These timescales are somewhat optimistic since the initial radius shrinks with time. However, it could be seen that initial larger radii lead to shorter migration/circularization timescales.

Initial larger radius shortens the migration and circularization timescales, which becomes ten times shorter in this case, as expected from the strong dependence of the tidal forces on the radius. 
The radius contacts to a final radius of $\gtrsim 1\ R_J$ within a Hubble time. A significant part of the migration takes place with a radius larger than $R_J$, since the migration timescale is shorter than the contraction timescale, manifesting the key role played by the initial inflated radius on the evolution. It should be noted that the radius of the initially non-inflated planet changes as well, but negligibly. 
In addition, the effective temperature is roughly constant at the early stages of the evolution, but increases as the planet gets closer to the star, since the irradiation from the star determines the effective equilibrium temperature of the planet.

\begin{figure*}
\centering
\includegraphics[width=\linewidth]{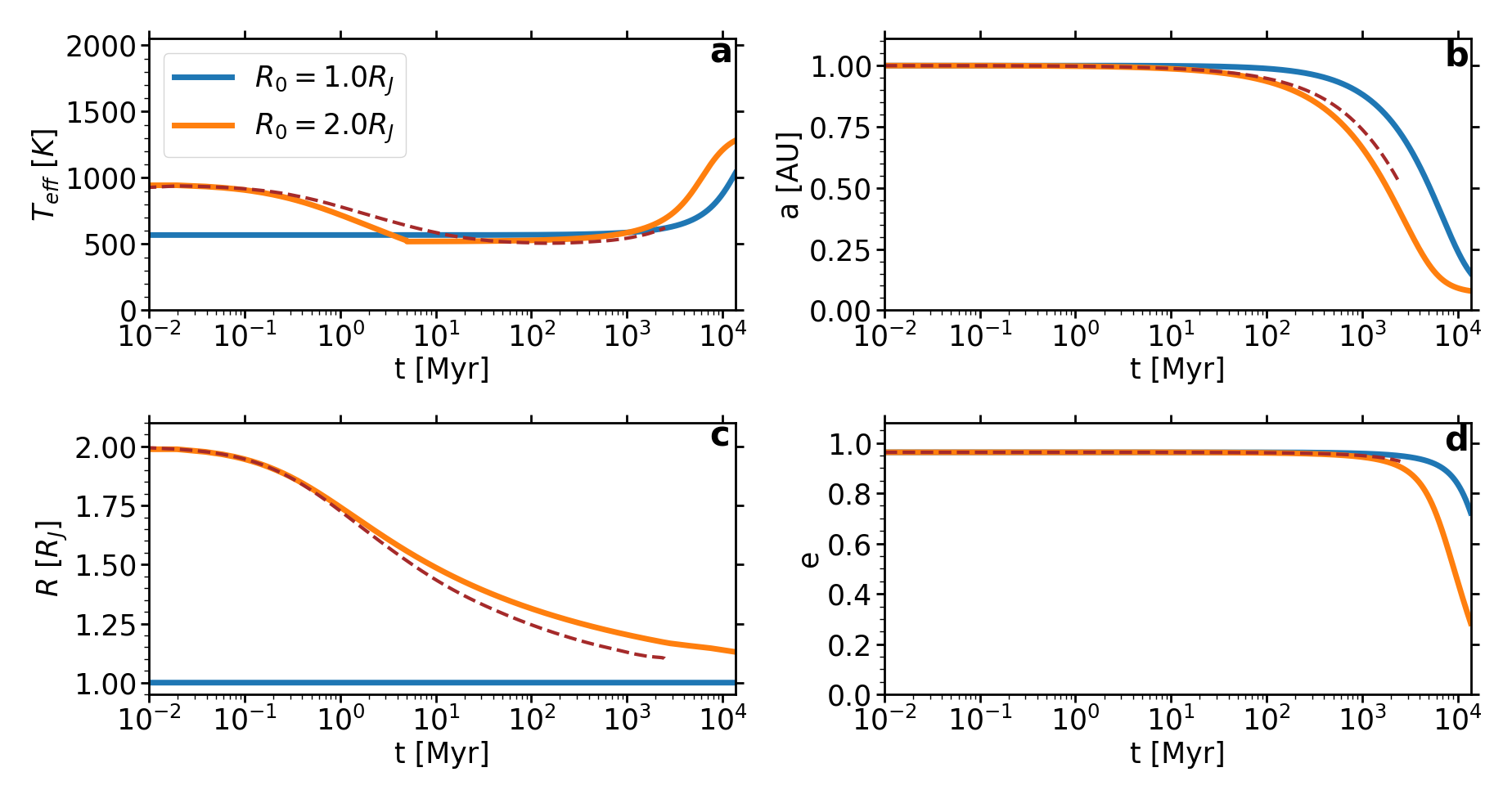}
\caption{
The thermal and orbital evolution of hot and warm Jupiter candidates migrating due to weak tides, including irradiation, for initial semimajor axis of $1 \ \rm{AU}$, initial radii of $1 \ R_J$ and $2 \ R_J$ and initial eccentricity of $0.963$ (blue -- $R_0 = 1 \ R_J$, orange -- $R_0=2 \ R_J$). The initial $2 \ \rm{R_J}$ model finalizes with an orbital period of $\sim 9.8 \ \rm{days}$ and the constant $1 \rm{R_J}$ with $24 \ \rm{days}$. The solid lines correspond to semi-analytical calculation and the dashed to the numerical. 
\textbf{a-d}, Time evolution of the effective temperature, semimajor axis, eccentricity and radius.}
\label{fig:WJ}
\end{figure*}

 In Fig. \ref{fig:WJ} we present the coupled thermal-dynamical evolution of a formed WJ and a HJ.
 Similar to Fig. \ref{fig:HJ}, the migration timescales are significantly shortened for a planet with the same given initial separation, eccentricity and mass but different initial radius. 
 
 We can see a good agreement between the semi-analytical model and the numerical one, throughout the evolution. Note that the numerical model terminates due to limitation of the used (older) \texttt{MESA} version in this regime; see more details on the termination criteria of the numerical model in paper II. In such cases, the numerical model could be extrapolated using the semi-analytical model. 
 
 We find that all the WJs produced through inflated eccentric migration though somewhat circularized are all still eccentric, and their final state is dictated by the initial angular-momentum and the energy dissipation rate. Inflated eccentric migration enhances the migration rate such that planets that would have otherwise never had become HJ or WJs when considering an initial and constant $1 \ R_J$ radii, migrate more efficiently and now WJs could become HJs and little/non migrating planets become WJs. Furthermore, inflated WJs, given the same initial conditions, would be less eccentric since they proceed faster in their migration; some of the expected WJs from the $1 \ R_J$ case become HJs once inflated migration is accounted for. This transitions between regimes, the effective accelerated change in the migration flow of gas-giants is discussed in more details in Sec. \ref{sec:Monte-Carlo}. 
 
 The semi-analytical approach provides an efficient, simple and computationally inexpensive approach to model the evolution, allowing us to consider a large phase space of initial conditions and consider a detailed population synthesis study that describes the evolution and formation of Jupiters. Nevertheless, a-priori, this simple approach can not describe the detailed internal structure of the planet, and might give rise to inaccurate modeling of the macroscopic properties of the planet and its evolution. 
 However, we generally find an excellent agreement between the numerical approach and the semi-analytical one, which therefore allows us to use the simple semi-analytical approach robustly. 

\section{Population synthesis study and occurrence rate estimates}\label{sec:Monte-Carlo}

In order to calculate the occurrence rate of HJs\&WJs and their properties, we use a population synthesis study.
Given the excellent agreement we found between the semi-analytic model and the numerical model (see Figs. \ref{fig:HJ},\ref{fig:WJ} and a further discussion in paper II), we rely on the analytic model as to provide fast an efficient evolution model, enabling us to study and analyze a wide range of initial conditions, and explore the formed population of HJs and WJs through a population synthesis study, without the need of the computational costly numerical simulation. 

We consider various plausible choices for the initial conditions of potential HJ/WJ progenitors (see Table \ref{table:rates_with_star_formation}), and randomly sample the parameter space of initial conditions. 

In order to characterize the population of HJs and WJs formed through regular eccentric migration and inflated eccentric migration, we make several assumptions regarding the initial conditions. We also reiterate that in this study we only consider eccentric migration following planet-planet scattering, where other secular evolution models for eccentricity excitation will be explored elsewhere. 

We assume a continuous star-formation rate in the Galactic disk, since the planet formation rate is proportional to the star formation rate (e.g. \citealp{Behroozi2015}) and therefore sample the age (i.e. evolution time) for each planet from a uniform distribution in the range $1-12\ \rm{Gyrs}$. We then evolve each planet in our sample using our semi-analytic approach, and examine its properties after a given time.
We define a HJ as a gas-giant with a final period shorter than $10 \ \rm{days}$, a define a (migrating) WJ as a gas-giant with a final period of $10-200\ \rm{days}$ and consider only cases where the semimajor axis shrank by at least factor of $2$ relative to the initial one as not to consider possible contributions from potential in-situ formed WJs. A gas-giant is assumed to be disrupted if its pericenter is smaller than the Roche radius. 
We define $f_{\rm HJ/WJ}$ as the fraction of formed HJs/WJs, as described above and use it to derive the total expected frequency of HJ\&WJ systems in the Galaxy. The fraction of formed HJs\&WJs for any given time $t$ is given by 

\begin{align}
f_{\rm HJ/WJ}(t)=\int \frac{dN}{dm_p}\frac{dN}{de_0}\frac{dN}{da_0}\frac{dN}{dR_{p,0}}\times \\ \nonumber
\times\chi(m_p,e_0,R_{p,0},a_0,t)de_0 da_0 dR_{p,0}dm_p 
\end{align}
where $dN/de_0, \ dN/da_0$ and $dN/dR_{p,0}$ are the differential distributions of the initial eccentricities, semimajor axes and initial planetary radii, respectively. $\chi$ is an indicator function determining whether a given system evolved to become a HJ/WJ after a time $t$, based on the semi-analytical coupled orbital-thermal evolution.
Calculating the fraction of formed HJs/WJs, we take into consideration also the possibility of tidal disruption and exclude the disrupted population from the formed HJs/WJs population. The tidal disruption radius is given by $r_{\rm dis}=\eta R_{p}\left(M_\star/M_p\right)^{1/3}$ where $\eta=2.7$ \cite{Guillochon2011}, and planets with pericenter approaching these value are assumed to be disrupted and, naturally,  are not considered to be HJ/WJs. Planets with a sufficiently large pericenter approach, that are negligibly affected by tides, and do not migrate, are not considered as migration-formed HJs/WJs, nor those that do migrate due to inflated eccentric migration, but do not migrate close enough to the star as to be considered as HJs or WJs in the relevant time considered.

In order to calculate the total occurrence rate, the frequency of HJs\&WJs among stellar systems is given by
\begin{align}\label{eq:rates}
\mathcal F_{\rm HJ/WJ} = f_J \times f_{2J} \times  f_{\rm unstable} \times f_{ecc,J} \times f_{\rm HJ/WJ}
\end{align}
where $f_J$ is the fraction of stellar systems hosting gas-giants,  $f_{2J}$ is the probability to find at least two gas-giants (in most of the cases, in order to scatter a gas-giant into a highly eccentric orbit leading to eccentric migration, another planet as massive as Jupiter is needed), $f_{\rm unstable}$ is the fraction of unstable systems, in which planet-planet scattering is likely to occur
and $f_{\rm ecc,J}$
is the fraction of sufficiently eccentric gas-giants that would evolve through eccentric migration to become HJs\&WJs. 

We study each choice of parameters (see the parameter sampling description below) for the initial distribution for $\gtrsim 10^4$ semi-analytical
simulations to determine the fractions of case which successfully evolve to become a HJ/WJ. Calculating this at at given time since birth (up to a chosen specific time), provides us the delay-time distribution, i.e. the fractions of HJ or WJ as a function of time since planet formation. As time goes by, planets that were observed as WJs could migrate further and become HJs; HJs could be disrupted and planets that were too far to be WJs could become ones. 
The rate in which the different areas in the parameter space are filled is determined by the initial conditions of the planet ($a_0,e_0,R_0$ and $m_p$) as well as the tide model (weak, dynamical etc.) but also on external energy sources that could slow the contraction.

We consider a specific star-formation history, and integrate the delay-time distribution (our Green function) weighted by the star-formation rate to obtain a realistic estimate for the current fractions of HJs/WJs in the Galaxy at current time, containing both young and old planetary systems. For disk stars, most relevant for currently observed exoplanet hosts, we consider a continuous, uniform rate of star-formation for the Galactic disc, as mentioned above. This is generally consistent with the inferred local star-formation history of the Galactic disc stellar populations where most exoplanets were observed to date. Note, however, that current exoplanets samples are dominated by those identified by the Kepler mission. The age distribution for Kepler stars peaks around $2.5 \ \rm Gyr$ and gradually falls off to larger ages \citep{Berger2020}. 

In the following we describe the parameters characterizing the initial conditions, the specific ranges of these parameters and their motivations. In order to evolve a gas-giant we need to ascribe a planet with both physical and orbital properties, including mass, initial radius, initial separation and initial eccentricity. 

\subsection{The Parameter Space of Initial Conditions}

Taking the observationally inferred occurrence rate and distributions of gas-giants as initial conditions is not fully self-consistent, since the observed systems had already been potentially affected by evolution. However, given the overall very low fractions of HJs\&WJs among the entire exoplanets population, and with the lack of direct data on the initial state of systems, it is likely that most systems did not significantly evolve after their initial formation, following the dissipation of the disk. Thus, the overall currently observed period and mass distribution are assumed to still reflects the post-formation initial conditions of gas-giants.

In the following we first discuss our choices for the initial distributions, and then present the resulting populations in the next section.

\textbf{Planetary radii:}
Planet formation models suggests that gas-giants form with large radii and contract rapidly to a radius of $\sim 4 \ R_J$, regardless of their mass, where a phase transition occurs and the gas-giant contracts and cools in a slower rate \citep{Guillot1996}. Therefore, we assume that the initial planet radii distribute uniformly between some lower and maximal radii, generally extending between $1 \ R_J$ and $4 \ R_J$, where we consider four possible sub-ranges, defined by $R_1-R_4$ (see Table \ref{table:rates_with_star_formation}). It should be noted that a more self-consistent choice of radii distribution should be taken from a planet-planet scattering simulation coupled to the radius evolution. This is out of the scope of the current paper and will be left out for future studies.

\textbf{Stellar and planetary masses:}
In this study we only consider only Sun-like stellar hosts all having the same (Solar) mass. 

The planetary masses are chosen from a power-law distribution with an exponent of $-1.1$, in the range $0.1-10$ M$_J$, consistent with observations \citep{Butler2006}. \\
\textbf{Semimajor axes:}
The semimajor axes of planets are assumed follow  a log-uniform distribution
and we consider the range between $0.4 \ \rm{AU}$ and $5 \ \rm{AU}$ (planets could be scattered into highly eccentric, with eccentricities close to $1$ orbits by $\sim 1 \ M_J$ planets for separations greater than $0.4 \ \rm{AU}$, \citealp{Dawson2018}). We also considered other distributions extending to larger separations, as mentioned below

\textbf{Eccentricities:}
The eccentricities are assumed to follow a Rayleigh distribution, i.e. $dN/de= \left(e/\sigma_e^2\right)\exp\left(-e^2/(2\sigma_e^2)\right)$, consistent with planet-planet scattering models (e.g. \cite{JuricTremaine2008} and references therein).
In order to shorten the running times, we sample the eccentricities starting from $0.8$ and then normalize properly the results according to the Rayleigh distribution.

\subsection{Normalization factors}
In the following we consider the various factors used to calculate the overall predicted occurrence rate of HJs\&WJs from inflated migration. \\
{\bf Gas-giant occurrence:} The occurrence rate of planetary systems hosting a gas-giant is of the order of $25\%$ \citep{Wang2015} but it depends on the metallicity, and could be as low as $5\%$ for low-metallicity hosts. 
Here we adopt a fiducial fraction of stellar systems hosting such planets to be $f_J=0.17$ for our model, to account for a non-extreme average case.
This fraction is decreased by considering only gas-giants that were not destroyed during the migration.
\\
{\bf Occurrence of planetary systems with at least two gas-giants:} We will assume that every system that has one gas-giant have at least two initially, i.e. $f_{2J}=1$ for a given $f_J$, generally consistent with observations \citep{Bry+16}, finding $>50\%$ of all planets residing between 1-5 AU have additional gas-giant companions, and considering that planet scattering typically eject most planets of the systems leaving behind 2-3 planets. 
\\
{\bf Fraction of (initially) unstable systems:} We take the fraction of unstable systems (given that they host at least two gas-giants), to be $f_{\text{unstable}}=0.75$. The actual number is unknown, however the overall consistency of the observed eccentricity distribution with a Rayleigh distribution suggest that planet-planet scattering due to unstable systems is ubiquitous \citep{FordRasio2008}, given that planets are generally thought to initially form on circular orbits.  
\\
{\bf Overall normalization pre-factor:}
Taken together, following Eq. \ref{eq:rates}, we get a normalization factor, $f_{\rm norm}=0.1257$, of $F_{\rm HJ/WJ}=f_{\rm norm}f_{\rm HJ/WJ}$, with $f_{\rm HJ/WJ}$ calculated from our models.

\section{Results}\label{sec:reults}
Our main results for the HJ/WJ occurrence rates from the various models considered are summarized in Table \ref{table:rates_with_star_formation}, in which we explore a large parameter space. We will note briefly the different parameters taken into consideration there: 
$\sigma_1=0.5, \ \sigma_2=0.4$, $R_1=1 \ R_J, \ R_2= 2\ R_J, R_3\sim U[2,4] \ R_J, \ R_4= 4 \ R_J
    $, $a_1\sim LU[0.4,5] \rm {AU}$, $d$ relates to dynamical tides model, $w$ to weak tides model, $c_y$ to deposition of $y\%$ of the irradiation at the center and tidal heating (corresponds to the tide model), such that the absence of $c_x$ stands for $0\%$ and $f_{x}$ to a variation in $f_{\rm dyn}$, i.e. $f_x$ stands for $f_{\rm dyn}=x$.
    $U$ stands for uniform distribution and $LU$ for loguinform distribution. 

The distributions of the orbital properties are shown only for our fiducial model, i.e. $\sigma_1R_2 a_1 df_{0.1}$.

We considered a variety of models which differ in the initial radii distribution of the planets, the initial eccentricity distribution and initial semimajor axes distribution. In addition we also consider different types of tidal evolution, either weak or dynamical, and different efficiencies of external heating (fraction of heat injected to the center of the planet, which depend on the not yet understood heat transfer processes).

\subsection{Occurrence Rates}

\begin{table*}
\setlength{\tabcolsep}{25pt}
\renewcommand{\arraystretch}{1.5}
    \centering
    \begin{tabular}{c||c|c | c| c|c}
     model & $\bf f_{\rm HJ}$& $ \bf f_{\rm WJ}$ & model & $\bf f_{\rm HJ}$& $ \bf f_{\rm WJ}$
      \\ 
     \hline
     \hline
              $\bf \sigma_1 R_1 a_1 w$ & 1\% & 0.3\%  &                  $\bf \sigma_1 R_1 a_1 d f_{0.1}$
          &
          2\% & 0.4\%
          \\
         $\bf \sigma_1 R_2 a_1 w$
        & 0.5\% & 0.4\% &
        $\bf \sigma_1 R_3 a_1 d f_{0.1}$ &
          1.6\% & 1\%
        \\
         $\bf \sigma_1 R_3 a_1 w$ &  0.4\%& 0.4\% &
                  $\bf \sigma_1 R_1 a_1 d f_{0.01}$ &  1.5\% & 0.2\% 
         \\
          $\bf \sigma_1 R_4 a_1 w$ & 0.2\% & 0.4\%
          &
                    $\bf\sigma_1 R_2 a_1 df_{0.01}$ &  1.2\%& 0.3\% 
          \\
          $\bf \sigma_2 R_1 a_1 w$ &  0.5\% &0.2\%
          &
                   $\bf \sigma_1 R_3 a_1 df_{0.01}$ & 1\% &0.4\% 
          \\
          $\bf \sigma_2 R_2 a_1 w$& 0.3\%
          & 0.2\%
          &
                   $\bf \sigma_1 R_4 a_1 df_{0.01}$ & 0.9\% & 0.5\%
          \\
                   $\bf \sigma_2 R_3 a_1 w$ & 0.2\% & 0.2\% &
                           $\bf \sigma_2 R_1 a_1 df_{0.01}$ & 0.8\% & 0.1\% 
                   \\
                   $\bf \sigma_2 R_4 a_1 w$ &  0.1\%&0.3\% &
                            $\bf \sigma_2 R_2 a_1 df_{0.01}$ & 0.7\% & 0.2\% 
         \\
                     $\bf \sigma_1 R_1 a_1 d f_1$ &
           2.5\% & 0.8\%
                                  &
                                  $\bf \sigma_2 R_3 a_1 df_{0.01}$ & 0.6\% & 0.3\% 
         \\
                              $\bf \sigma_1 R_3 a_1 d f_1$ & 1.8\% & 2\% 
                        
             &
                                               $\bf \sigma_2 R_4 a_1 df_{0.01}$ & 0.6\% & 0.4\%
     \\
              $\bf \sigma_1 R_1 a_1 d c_{1} f_{0.1}$
        & 0.1\% & 0.2\% 
         &
          $\bf \sigma_1 R_3 a_1 d c_{1}f_{0.01}$ & 0.2\% & 0.6\%
         \\
             $\bf \sigma_1 R_3a_1 dc_{10}f_{0.01}$ &
         $0.1\%$ & $0.6\%$
         &
                             $\bf\sigma_1 R_1a_1 dc_{10}f_{0.01}$ &
         $0.1 \%$ & $0.2\%$
        \\
        $\bf \sigma_1 R_3 a_1 dc_1 f_{0.1}$ & 0.06\% & 0.7\% 
     \end{tabular}
    \caption{
    Normalized formation fractions of HJs and WJs -- $f_{\rm HJ}$ and $f_{\rm WJ}$ correspondingly, as derived from the Monte-Carlo simulation, based on the semi-analytic model.
    $\sigma_1=0.5, \ \sigma_2=0.4$, $R_1=1 \ R_J, \ R_2= 2\ R_J, R_3\sim U[2,4] \ R_J, \ R_4= 4 \ R_J
    $, $a_1\sim LU[0.4,5] \rm {AU}$, $d$ relates to dynamical tides model, $w$ to weak tides model, $c_x$ to deposition of $x\%$ of the irradiation at the center and tidal heating (corresponds to the tide model), such that the absence of $c_x$ stands for $0\%$ and $f_{x}$ to a variation in $f_{\rm dyn}$, i.e. $f_x$ stands for $f_{\rm dyn}=x$. The results are based on a statistics from $\geq 10^4$ Monte-Carlo simulations per each set of initial parameters. \label{table:rates_with_star_formation}
    }
\end{table*}
As can be seen in table  \ref{table:rates_with_star_formation}, the different models and physical processes included give rise to large differences in fractions of HJs and WJs up to factors of 10-20 between the most extreme cases. The models we consider cannot robustly reproduce the the observationally inferred occurrence rates of HJs ($F^{\rm Obs}_{\rm HJ}=0.3-1.5\%$). However, our strong dynamical tides models (df1 and df0.1), without efficient central heating give occurrence rates of 1.6-2.5$\times f_{\rm norm}=0.2-0.32\%$, i.e. marginally reproduce the lower estimates for the observed HJ occurrence rate. The same models can robustly reproduce the occurrence rates of eccentric WJs ($F^{\rm Obs}_{\rm WJ}=0.04-1.37\%$, given that $30-90\%$ of all WJs are assumed to have formed in-situ, or through disc migration, that produce low-eccentricity WJs; \citealp{HuangWuTriad2016}). We find occurrence rates of 0.4-2$\times f_{\rm norm}=0.05-0.26\%$, in our strong dynamical tides models. 
In general, we find that the consideration of initially inflated gas-giants give rise to 2-3 higher occurrence rates of WJs compared with models of constant non-inflated gas-giants, but decrease the occurrence rate of HJs by 20-30$\%$ and up to a factor of 2 in some cases.
 
The occurrence rate of HJs could be enhanced by the more rapid inflated eccentric migration, bringing them in from larger distances, compared with non-inflated planets. However, we find that the overall occurrence rate of HJs  decreases by inflated migration. This is due to the enhanced tidal disruption of the now larger gas-giants (and correspondingly larger Roche-radius) when they are first scattered into high eccentricities see Fig. \ref{fig:diagram}). 
The tidal disruption is determined merely by the initial conditions, apart from cases of extremely efficient heating that leads to re-inflation. Hence, decoupling in latter stages from the planet-planet scattering phase that is expected to lead to initial lower radii leads also to higher formation fraction of HJs.
In contrast, the occurrence rate of WJs increases, due to the flow in the parameter space that enables gas-giants that would have otherwise (if they weren't initially inflated) not migrated (or little migrated) to migrate more rapidly as to attain a sufficiently small semimajor axes as to become WJs, while the migration of WJs to become HJs does not increase at a similar level.  

\subsection{Parameter Space Evolution}

\begin{figure}
\centering
\includegraphics[width=\linewidth]{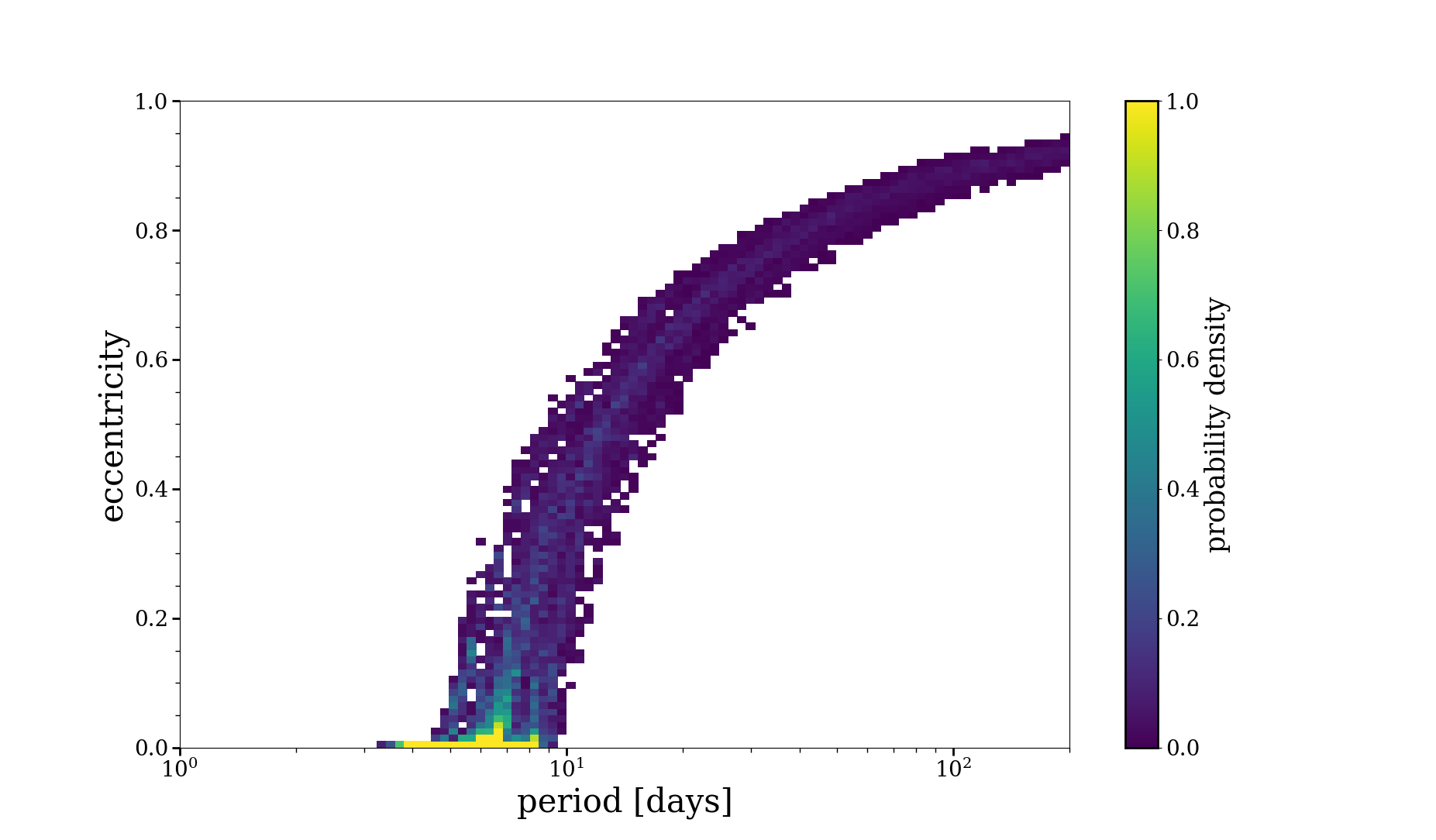}
\caption{ 
2-dimensional histogram of the eccentricity and orbital period of hot and warm Jupiters after a Hubble time, as obtained from the population-synthesis models initialized with $\sigma_1R_3 a_1df_{0.1}$. The probability density is normalized  according to the total fraction of successful formation of HJs and WJs among all initial conditions sampled. 
}
\label{fig:Monte Carlo density}
\end{figure}

Fig. \ref{fig:Monte Carlo density} shows the resulting distribution of the eccentricities and orbital period of our population-synthesis models initialized with $\sigma_1 R_3 a_1d f_{0.1}$, where we find the majority of HJs formed via the eccentric migration to have been circularized, whereas  the WJs are eccentric, suggesting that their migration process has not yet terminated. The empty region on the left reflects the tidal disruption of planets with pericenter below the Roche-Radius. The empty region on the right reflects the conservation of angular momentum along the migration, such that an initial high eccentricity sets a lower bound on the semimajor axis; planets with too-high per-centers are not affected by tides and do not migrate.

\begin{figure}
\centering
\includegraphics[width=1.15\linewidth]{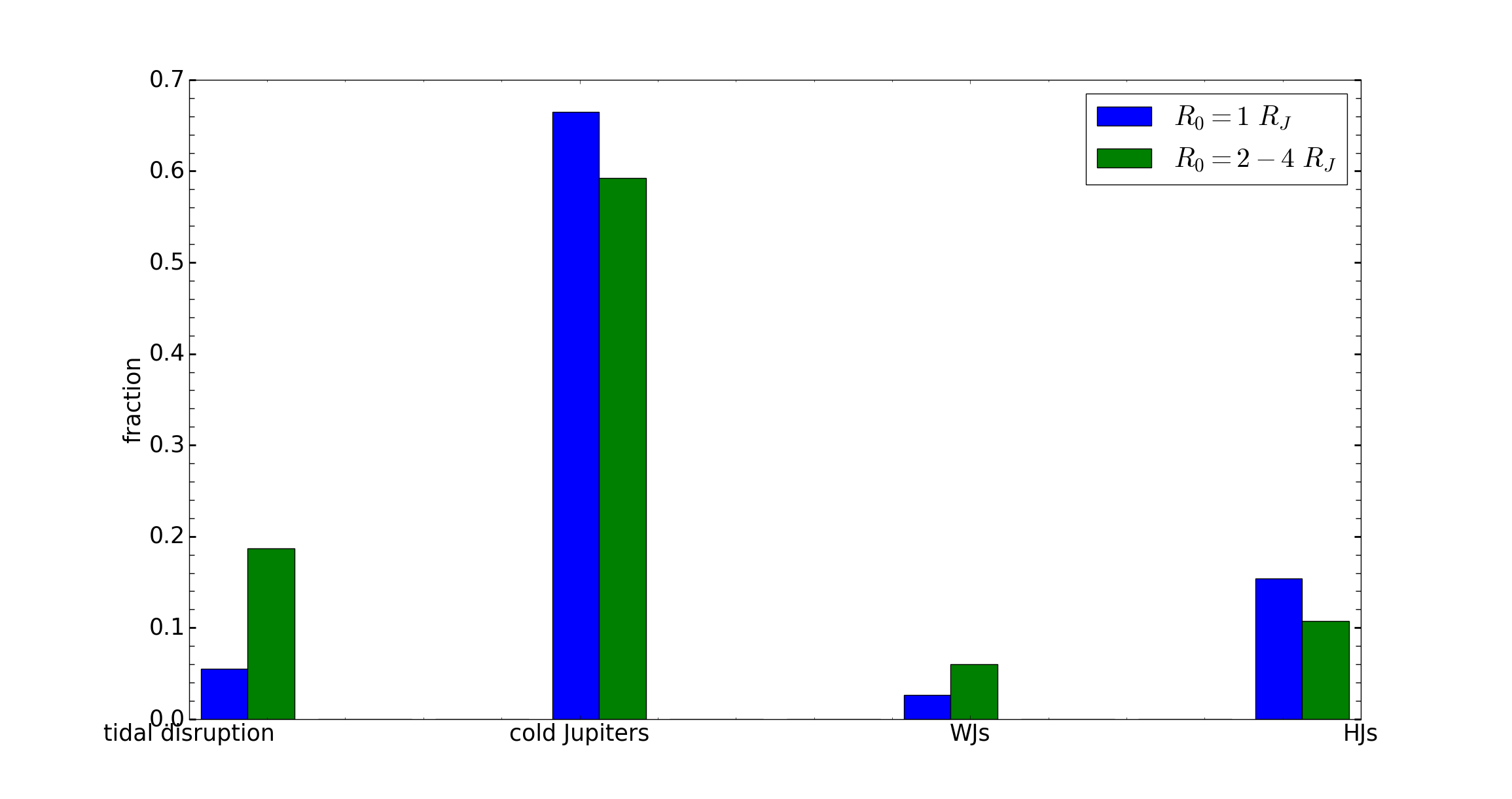}
\caption{ 
A diagram of the final fractions concluded from the Monte-Carlo simulation and a comparison inflated ($\sigma_1 R_3 a_1 d f_{0.1}$) and non-inflated ($\sigma_1 R_1a_1df_{0.1}$) initial radii. The fractions could be summed over to $1$ -- we present just the results of the simulation without further normalization. The results are based on $10^4$ runs of the semi-analytical model per each case.
}
\label{fig:diagram}
\end{figure}

In Fig. \ref{fig:diagram} we present a histogram describing the final fractions of the possible outcomes of eccentric inflated migration: cold Jupiters, WJs, HJs or tidal disruption. The population is dominated by cold Jupiters, i.e. gas-giants with periods larger than $200 \rm {days}$ or gas-giants that didn't migrate at least half of their initial semimajor axis. Inflated eccentric migration reduced the percentage of cold Jupiters, since the migration is more efficient in this model, which leads to a more significant flow in the parameters space from the cold Jupiters regime to the WJs. The fraction of WJs also increases due to inflated eccentric migration, from the similar considerations. However, the fraction of HJs is reduced, due to tidal disruption. 

\begin{figure}
\centering
\includegraphics[width=1.\linewidth]{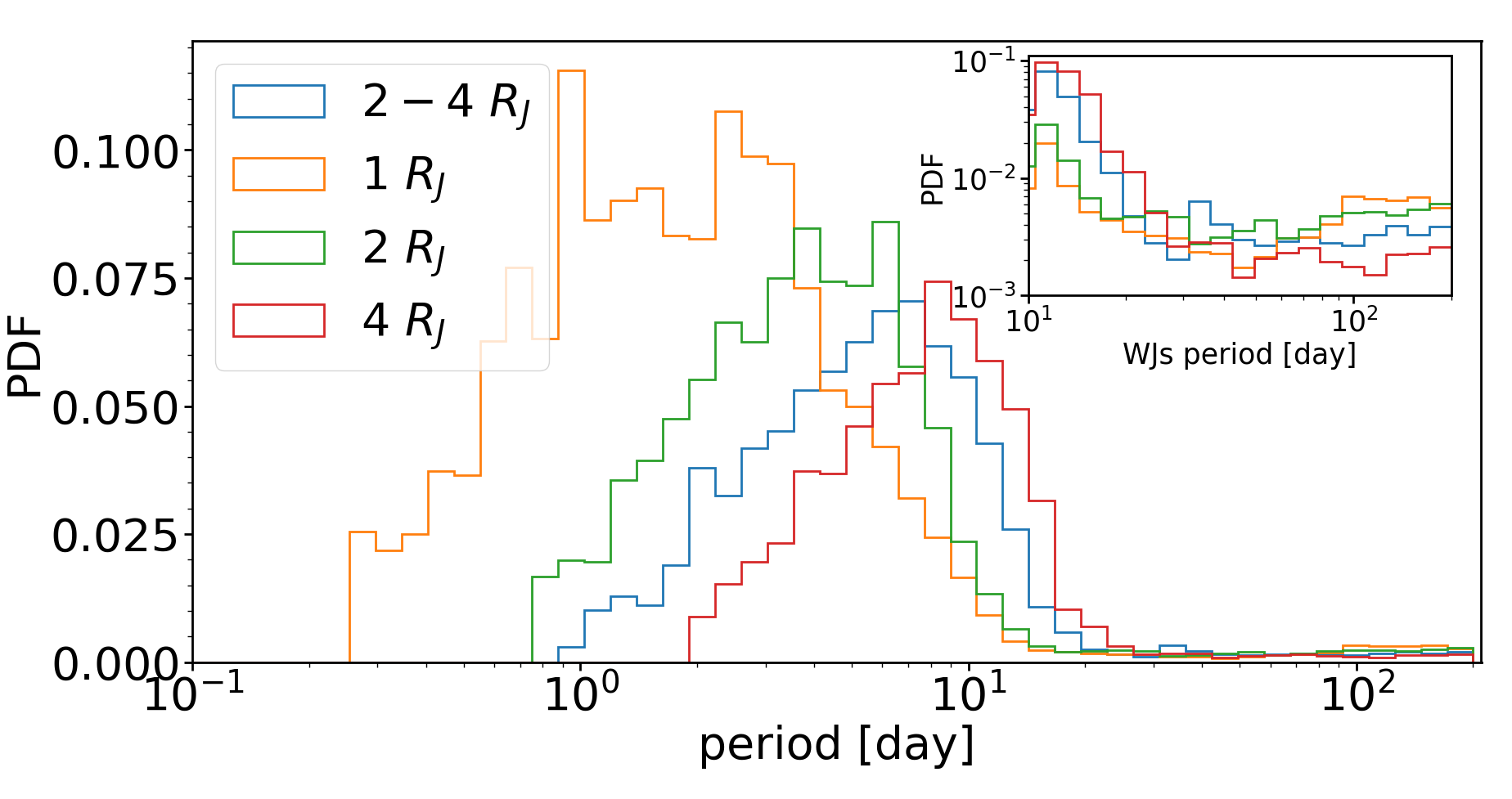}
\caption{ 
The period distribution as found by the Monte-Carlo simulation, based on the semi-analytical model, for different initial radius distribution. The rest of the parameters are drawn according to $\sigma_1R_3a_1df_{0.1}$. In the inset figure we introduce the probability distribution function of WJs only (with the same color code)}.
\label{fig:period distribution -- radius}
\end{figure}

\begin{figure}
\centering
\includegraphics[width=1.\linewidth]{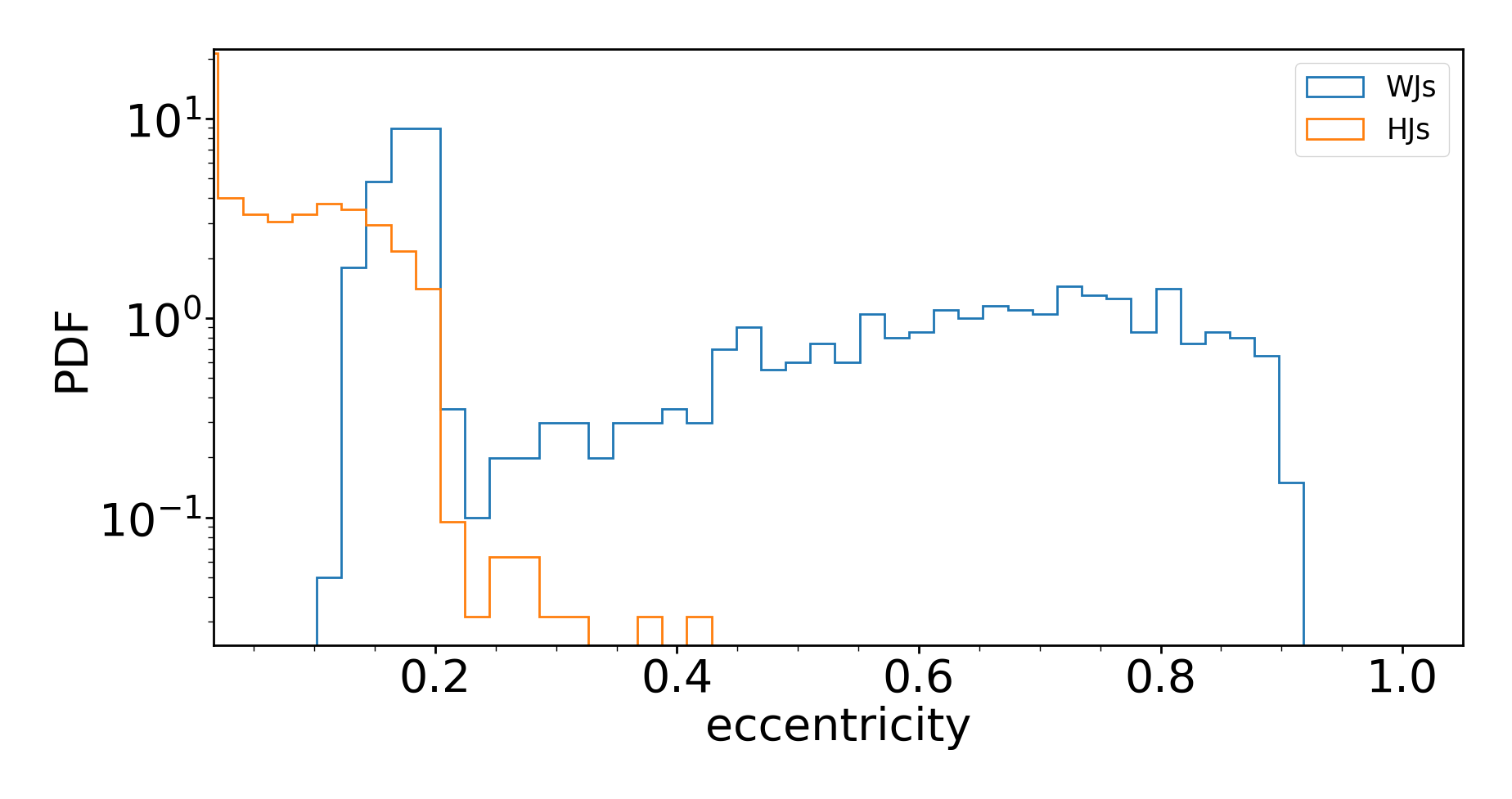}
\caption{ 
The eccentricity distribution of HJs and WJs as found by the Monte-Carlo simulation, based on the semi-analytical model, after $1 \ \rm{Gyr}$. The the parameters are drawn according to $\sigma_1R_3a_1df_{0.1}$.}
\label{fig:eccentricity}
\end{figure}

In Fig. \ref{fig:eccentricity} we present the eccentricity distribution of HJs and WJs. It could be seen that within this time, HJs tend to circularize, while WJs tend towards higher eccentricities, although could obtain lower eccentricities, with the lowest at $\lesssim 0.1$. 

\subsection{HJs \& WJs Population}
HJs\&WJs migrate faster via the inflated eccentric migration channel, as manifested also for a specific case in Fig. \ref{fig:HJ}.
The formed HJs population via inflated eccentric migration can be distinguished from non-inflated migrating giants. The distribution of the formed inflated HJs is shifted towards larger periods, as can be seen in Fig. \ref{fig:period distribution -- radius}, in agreement with the findings of \cite{Petrovich2015b}, that conducted a population study of cooling initially inflated giants (where no external heating considered in that study) for a secular evolution channel. The larger size of inflated HJs make them susceptible to tidal disruption at larger pericenter approaches, as can be seen directly from the expression for Roche limit $r_{\rm dis}=\eta R_{p}\left(M_\star/M_p\right)^{1/3}$, due to the increased radius and small pericenter. 

Indeed, the fraction of tidally disrupted planets significantly increases for models with inflated giants (e.g. see Fig. \ref{fig:diagram}). These disruptions arise already from the initial conditions, and not the later evolution. Planets are initiated at their largest sizes following their formation, and can therefore be disrupted at higher pericenter approaches at these times, but than the Roche-radius increases as the planets contract.   

Only in a cases where significant external tidal/radiative heating is able to re-inflate planets, do tidal disruptions occur during the migration. Indeed, models with significant heating of the central parts (c1, c10) show even higher tidal disruption rates, and, moreover, such disruptions occur during their migration, following their tidal and radiative inflation and not immediately after their initial scattering to high eccentricities.

Star formation modifies both the populations of HJs and WJs. As a convolution of a single-time star formation events, it gives rise to further formation of WJs, together with a smaller fraction of HJs and indicates that WJs might be younger than HJs, since gas-giants that are currently observed as WJs could migrate in and become HJs. 
We therefore predict that on average, WJs should reside in younger systems than HJs, if eccentric migration plays a significant role in their production.

\subsection{The Effects of External Heating on The Population}

External heating leads to a slowed cooling and hence increases both in in the production of WJs as well as increase the tidal disruptions rate (see e.g. table \ref{table:rates_with_star_formation} and Fig. \ref{fig:diagram}). In terms of the WJs-HJs parameter space, external heating speeds up the flow from cold Jupiters to WJs and so on; taken together it gives rise to an increased total number of WJs, compared with HJs and tidal disruptions.

The efficiency of external heating deposition depends both on the depth of the deposition, its duration and its amplitude. For example, irradiation deposited at the outer layers of the gas-giants contributes mainly to the effective temperature, but makes a negligible contribution to heating of the central parts of the planet, and consequently little affects the radius of the planet, nor the dynamical tidal evolution which strongly depends on the radius. 

This is not the case if a fraction of the irradiation is deposited at the center. In this case there could be a significant effect that might even lead to inflation, if the fraction of centrally deposited is sufficiently large. The exact process of energy transfer from the outer planets to the interior is still unknown and several mechanisms were suggested (e.g. \citealp{ArrasSocrates2010,BatyginStevenson2010,YoudinMitchell2010}), where the main motivation for these are the observations of old inflated HJs, which likely require such efficient heat transfer mechanisms.
We encapsulate the uncertainties in the external heating source and the depth of deposition in an efficiency of deposition at the center, similar to other studies, that focused on the thermal evolution rather than on the coupled thermal-dynamical evolution (e.g. \citealp{Bodenheimer2001,Komacek2020}). Our cetral heat deposition models therefore do not correspond speficialy to any of the suggested models, but rather bracket the potential effects of potential efficient heat transfer.

\subsection{Dependence on Parameters}

The final population and its properties depend strongly on the choice of initial distributions and their parameters. While some of the distributions are well-constrained from observations, others suffer from large uncertainties, which we account for by considering several choices of parameters. In the following we discuss several possible choices of the parameters and their effect on the final distributions. 

In Fig. \ref{fig:period distribution -- radius}, we present the dependence on the final periods on the initial radii distribution. The formed population via the inflated eccentric migration channel peaks in a larger period and gives rise to an enhanced filling of the available parameter space of WJs.  

\begin{figure}
\centering
\includegraphics[width=1.15\linewidth]{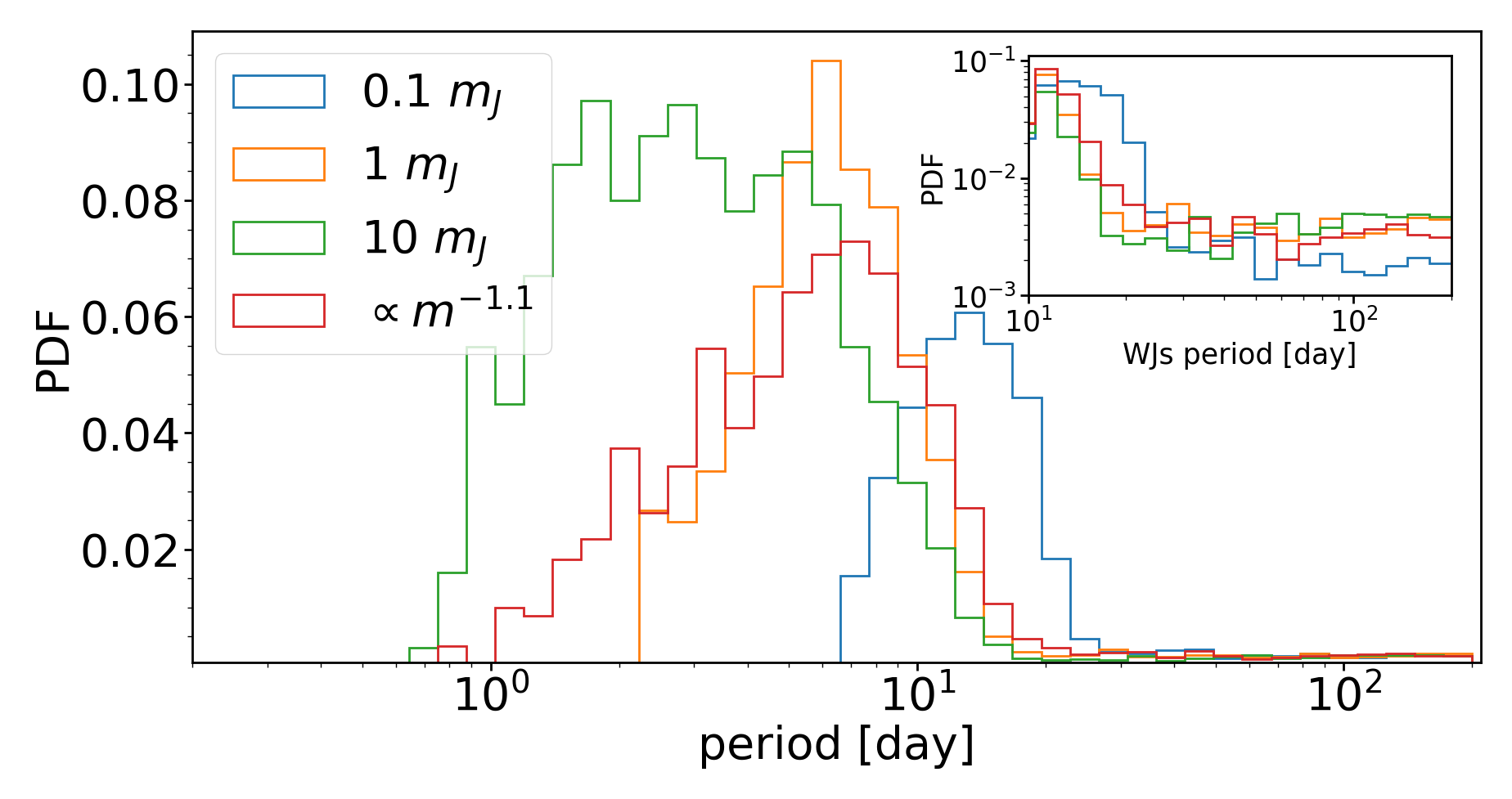}
\caption{ 
The period distribution as derived from the Monte-Carlo simulation, based on the semi-analytical model, for different masses, normalized to $1$. The rest of the parameters are drawn according to $\sigma_1 R_3 a_1 df_{0.1}$. The red line corresponds to our standard mass distribution-- $\propto m^{-1.1}$ within the range $[0.1,10] \ m_J$.In the inset figure we introduce the probability distribution function of WJs only (with the same color code)}
\label{fig:period distribution -- masses}
\end{figure}

As shown in Fig. \ref{fig:period distribution -- masses}, the initial population of more massive gas-giants gives rise to a post-migration population residing in smaller periods. This might be expected given that migration is more efficient for massive planets, as can be seen directly from the tidal evolution equations (e.g. eqns. \ref{eq:weak tide} and \ref{eq:dynamical tides} for weak and dynamical tides correspondingly), although the this is not trivial given the transitions and flow between cold-Jupiters, WJ, HJs and tidally disrupted planets which changes in each model.
It should be noted that the migration timescale of lower mass planets is shorter than more massive ones (see also in paper II).  However, they are more vulnerable to tidal disruption, such that the overall effect is that lower masses are more efficient in the production of WJs rater than HJs.

The observational mass distribution sets more weight on the less massive gas-giants, such that the total period distribution flattens to include a larger range of periods, from HJs to WJs. 

\begin{figure}
\centering
\includegraphics[width=1.\linewidth]{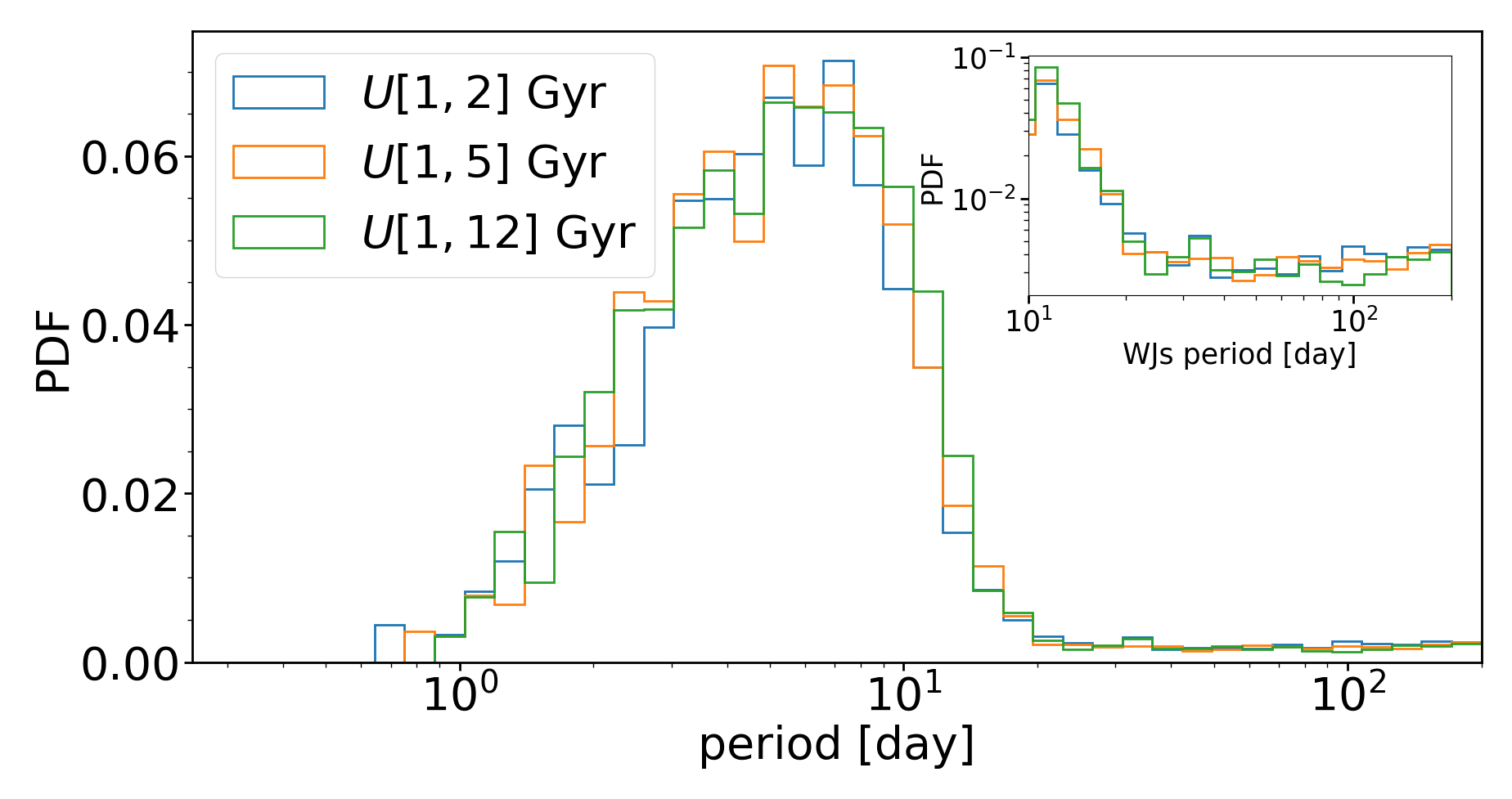}
\caption{ 
The period distribution as derived from the Monte-Carlo simulation, based on the semi-analytical model, after different times, normalized to $1$, considering continuous star formation. The rest of the parameters are drawn according $\sigma_1 R_3 a_1 df_{0.1}$.In the inset figure we introduce the probability distribution function of WJs only (with the same color code)}
\label{fig:period distribution -- times}
\end{figure}

\begin{figure}
\centering
\includegraphics[width=1.15\linewidth]{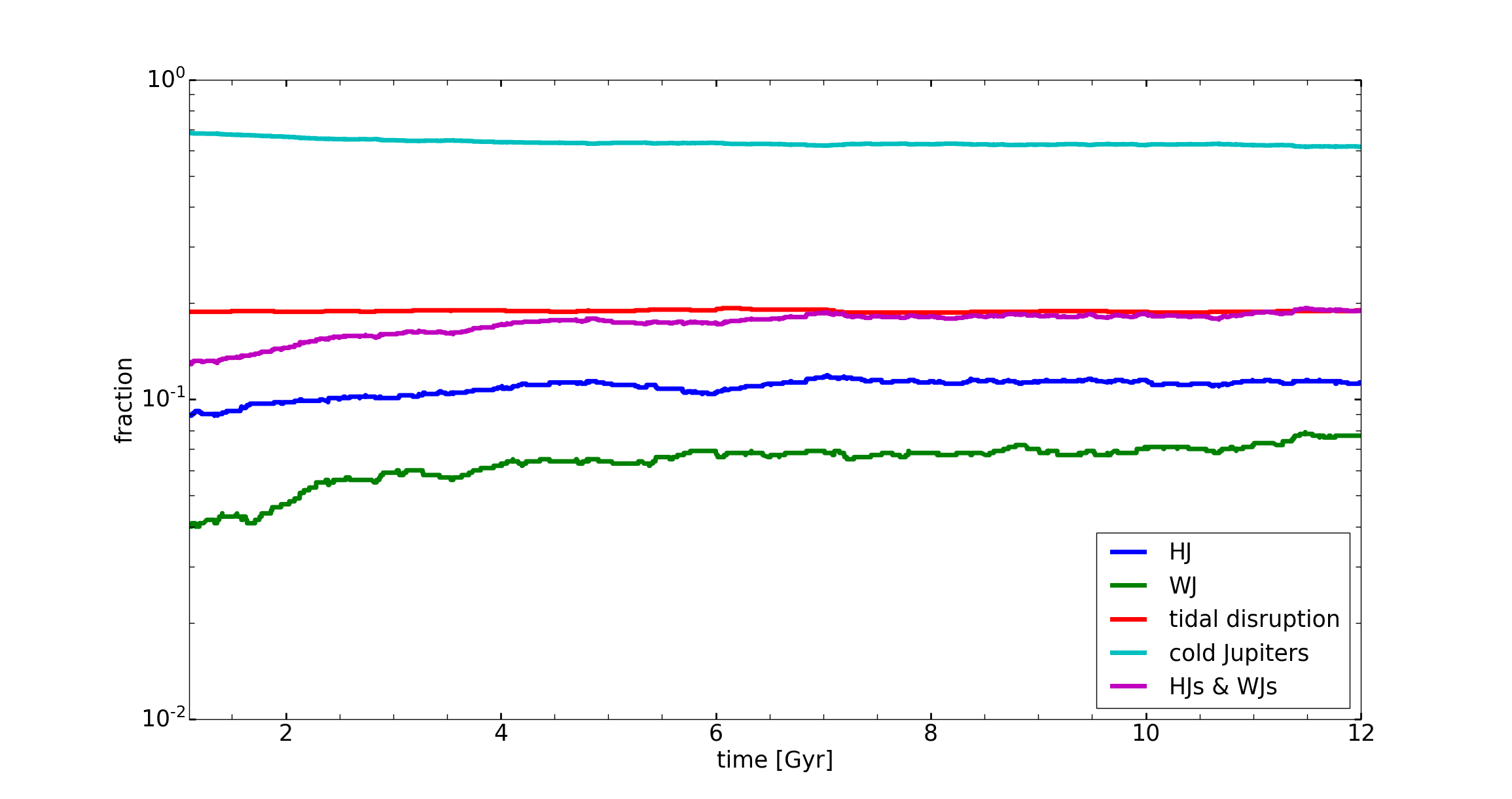}
\caption{ 
Delay time distribution, as derived from the Monte-Carlo simulation for $\sigma_1 R_3 a_1 d f_{0.1}$, after $1$ star formation event. The plot is normalized to $1$ (for every time) and is based on $10^4$ runs of the semi-analytical based Monte-Carlo per each case. 
}
\label{fig:delay time}
\end{figure}

\begin{figure}
\centering
\includegraphics[width=1.\linewidth]{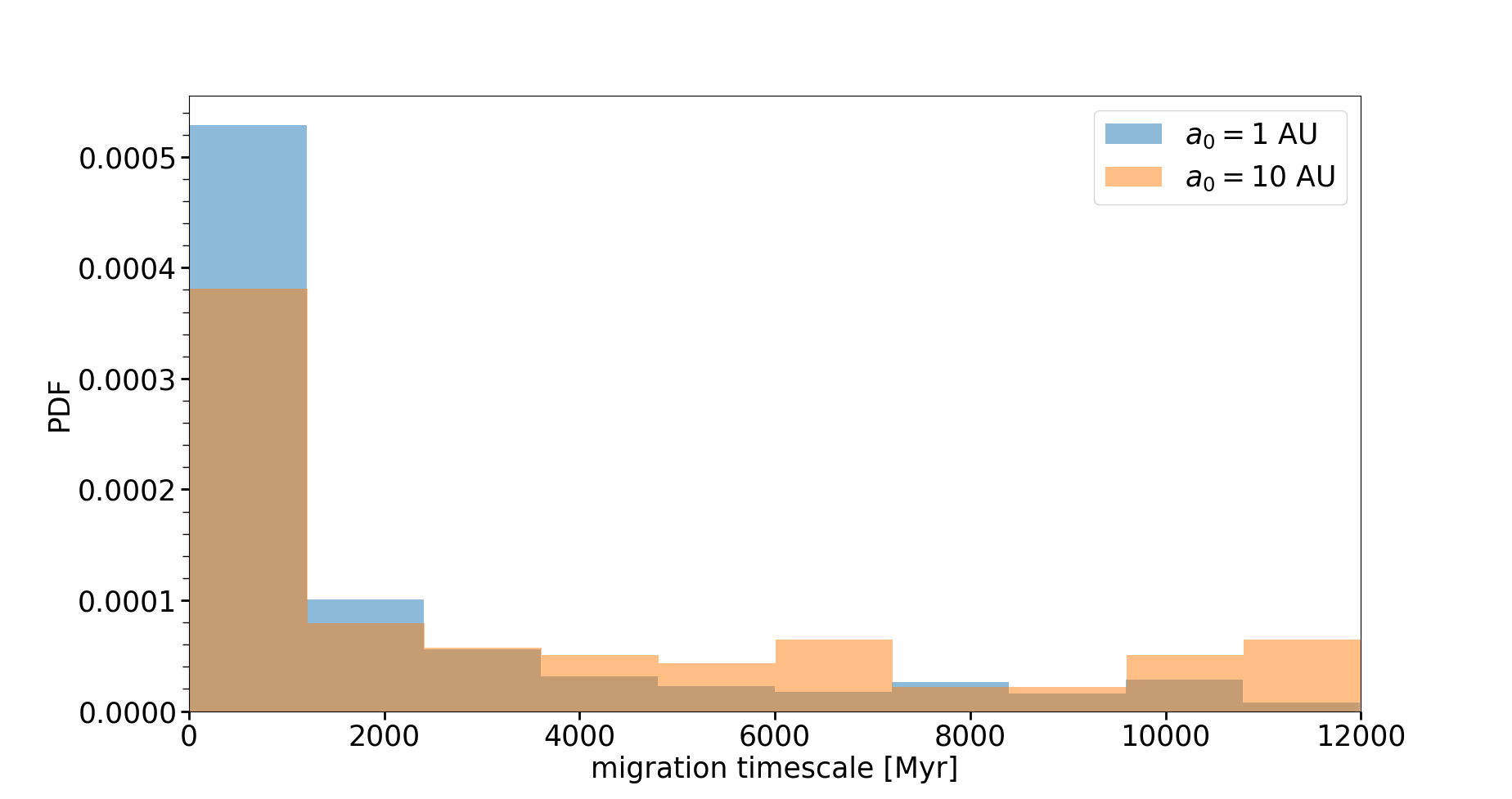}
\caption{Histogram of the migration timescales of HJs, for different initial semimajor axes, after $12 \ \rm{Gyr}$ from a single star formation event, as derived from our population synthesis based on the semi-analytic model (the rest of the parameters are sampled according to our fiducial model).
}
\label{fig:mig_time_a}
\end{figure}

\begin{figure}
\centering
\includegraphics[width=1.\linewidth]{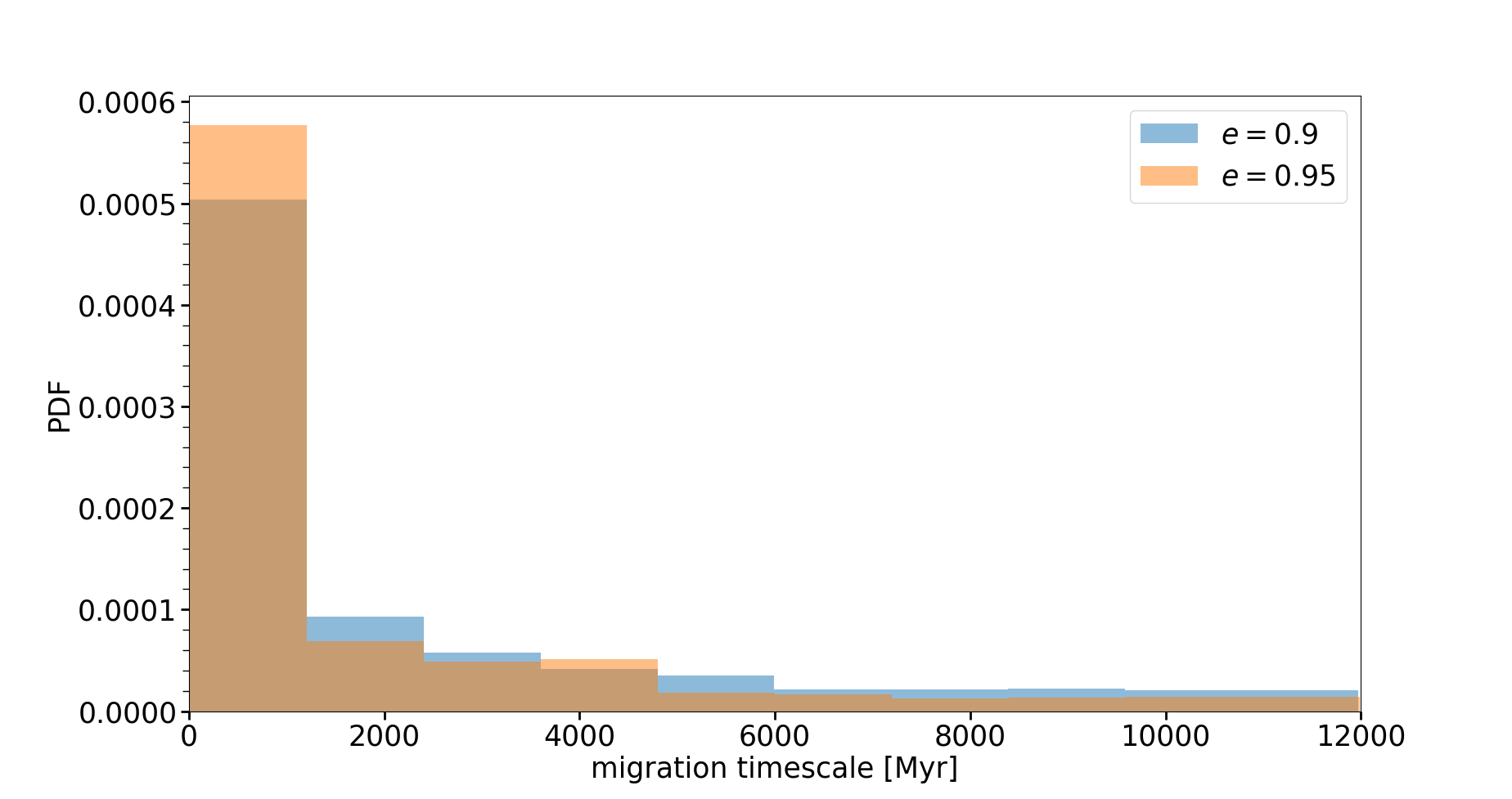}
\caption{Histogram of the migration timescales of HJs, for different initial eccentricities, after $12 \ \rm{Gyr}$ from a single star formation event, as derived from our population synthesis based on the semi-analytic model (the rest of the parameters are sampled according to our fiducial model).
}
\label{fig:mig_time_e}
\end{figure}

\begin{figure}
\centering
\includegraphics[width=1.\linewidth]{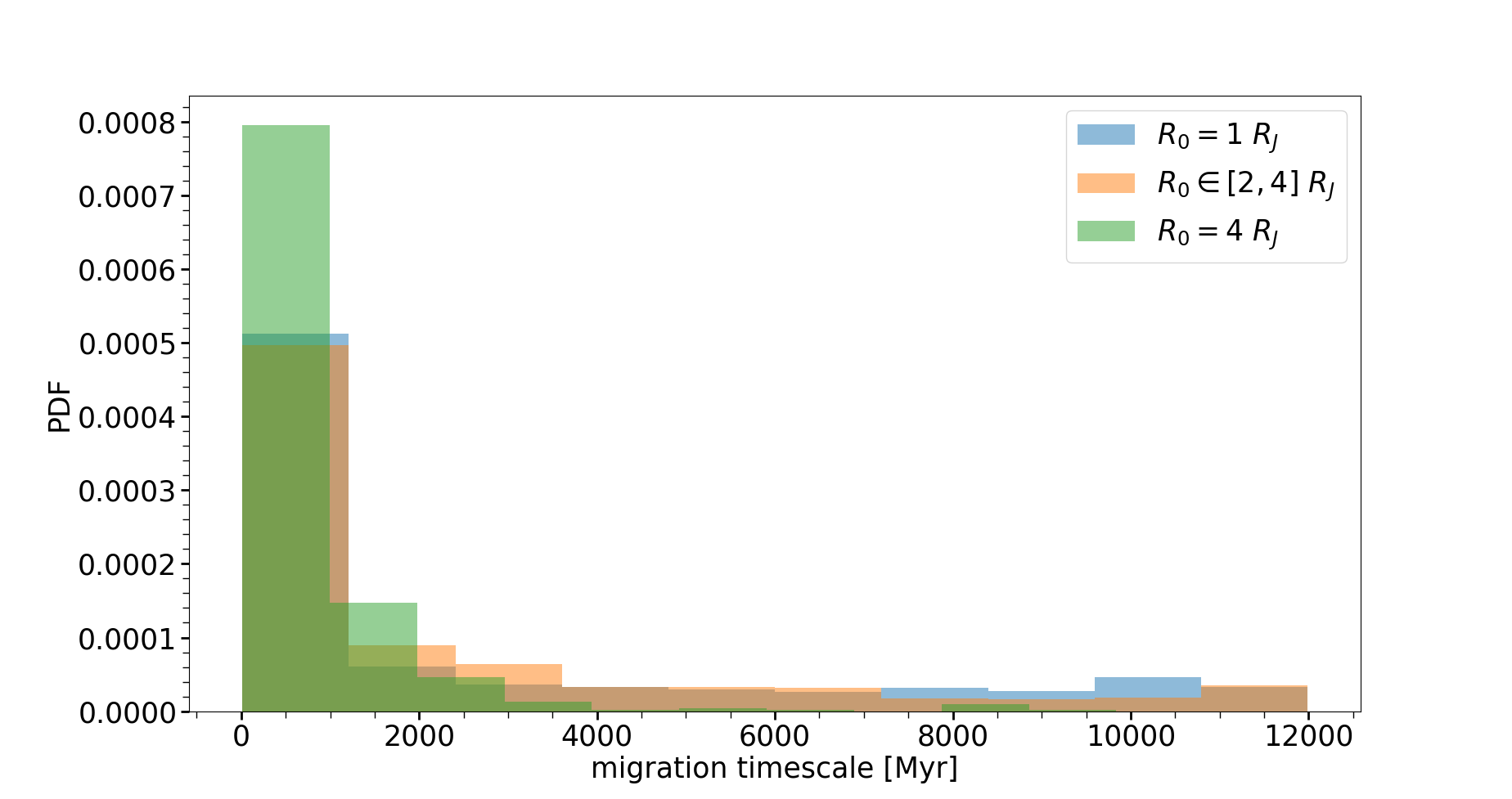}
\caption{Histogram of the migration timescales of HJs, for different initial radii, after $12 \ \rm{Gyr}$ from a single star formation event, as derived from our population synthesis based on the semi-analytic model (the rest of the parameters are sampled according to our fiducial model).
}
\label{fig:mig_time_R}
\end{figure}

In Figs. \ref{fig:period distribution -- times} and \ref{fig:delay time} we show the time evolution of the gas-giants population. As time goes by, more and more cold Jupiters migrate inwards to become WJs, some of them migrate to become HJs and some will be migrate further and be disrupted. It can be seen following the peak of the distribution. At early times, there is a fast rise in HJs which form more rapidly, but at later times, WJs form and the peak gradually moves towards larger periods. The vast majority of the HJs migrate via timescales shorter than $1 \ \rm{Gyr}$. Considering star formation leads to a continuous flow of formed HJs and could basically be understood as a time convolution of the distribution derived from a single star formation event. In models without central heating, planets do not reinflate, and all tidal disruptions occur promptly following planets scattering, while tidal and radiative heating do not inflate the planets which attain their maximal radii following their formation. When central heating is efficient planets can reinflate and become larger than their original radii and be more prone to tidal disruptions at later times. The exact amount of heat needed for reinflation could be roughly estimated by setting the total luminosity to be larger than $0$, as can be seen in eq. \ref{eq:dRdt}.
The effect of reinflation could be seen also from the decreased fraction obtained for gas giants with large amounts of energy injected in their center (see Table \ref{table:rates_with_star_formation}).

\begin{figure}
\centering
\includegraphics[width=1.15\linewidth]{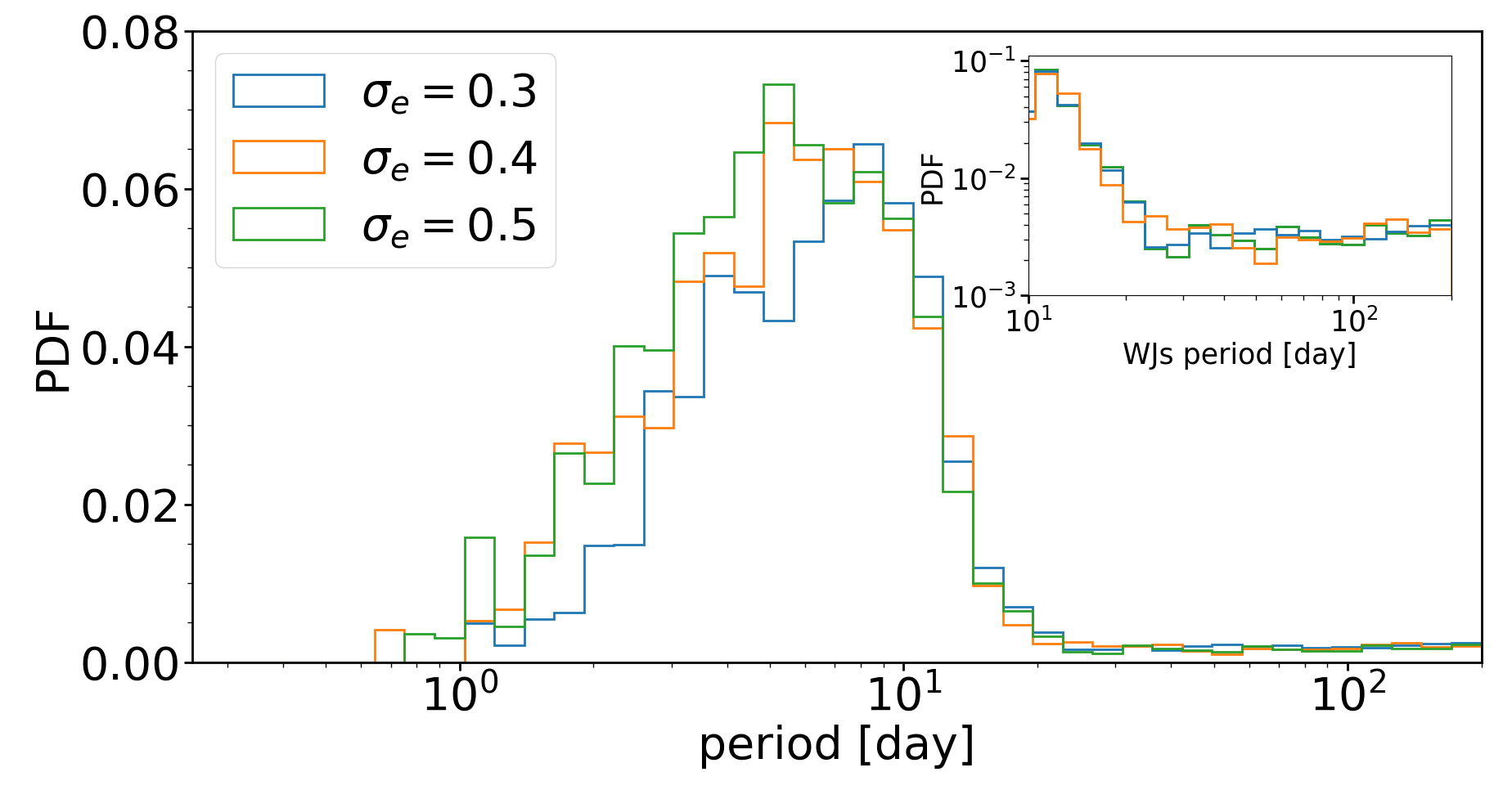}
\caption{ 
The period distribution as derived from the Monte-Carlo simulation, based on the semi-analytical model, for different masses, normalized to $1$. The rest of the parameters are drawn according to $\sigma_1 R_3 a_1 df_{0.1}$, that corresponds to $\sigma_e=0.5$. In the inset figure we introduce the probability distribution function of WJs only (with the same color code)}
\label{fig:period distribution -- eccentricity}
\end{figure}

In Figs. \ref{fig:mig_time_a}, \ref{fig:mig_time_e} and \ref{fig:mig_time_R} we present the dependence of the migration timescale on the initial semimajor axis and eccentricity correspondingly. As expected, larger initial semimajor axes lead to larger migration timescales of HJs, and higher initial eccentricities lead to more efficient migration that final leads to smaller migration timescales. Larger initial radii lead to extremely short migration timescales, while long timescales are cut of the histogram, due to the elevated disruption rates. 

In Fig. \ref{fig:period distribution -- eccentricity} we present the dependence on the dispersion of the eccentricity distribution.
Lower eccentricity distributions give rise to larger rates of WJs, but smaller rates of HJs, as expected. Eccentric tidal migration is more efficient, i.e. extracts more energy from the orbit, when larger eccentricities are included. Since the WJs produced via inflated migration could be thought as transient HJs that didn't manage to end their migration after a given time, their fraction increases when a lower eccentricity dispersion is taken into consideration. 
\begin{figure}
\centering
\includegraphics[width=1.\linewidth]{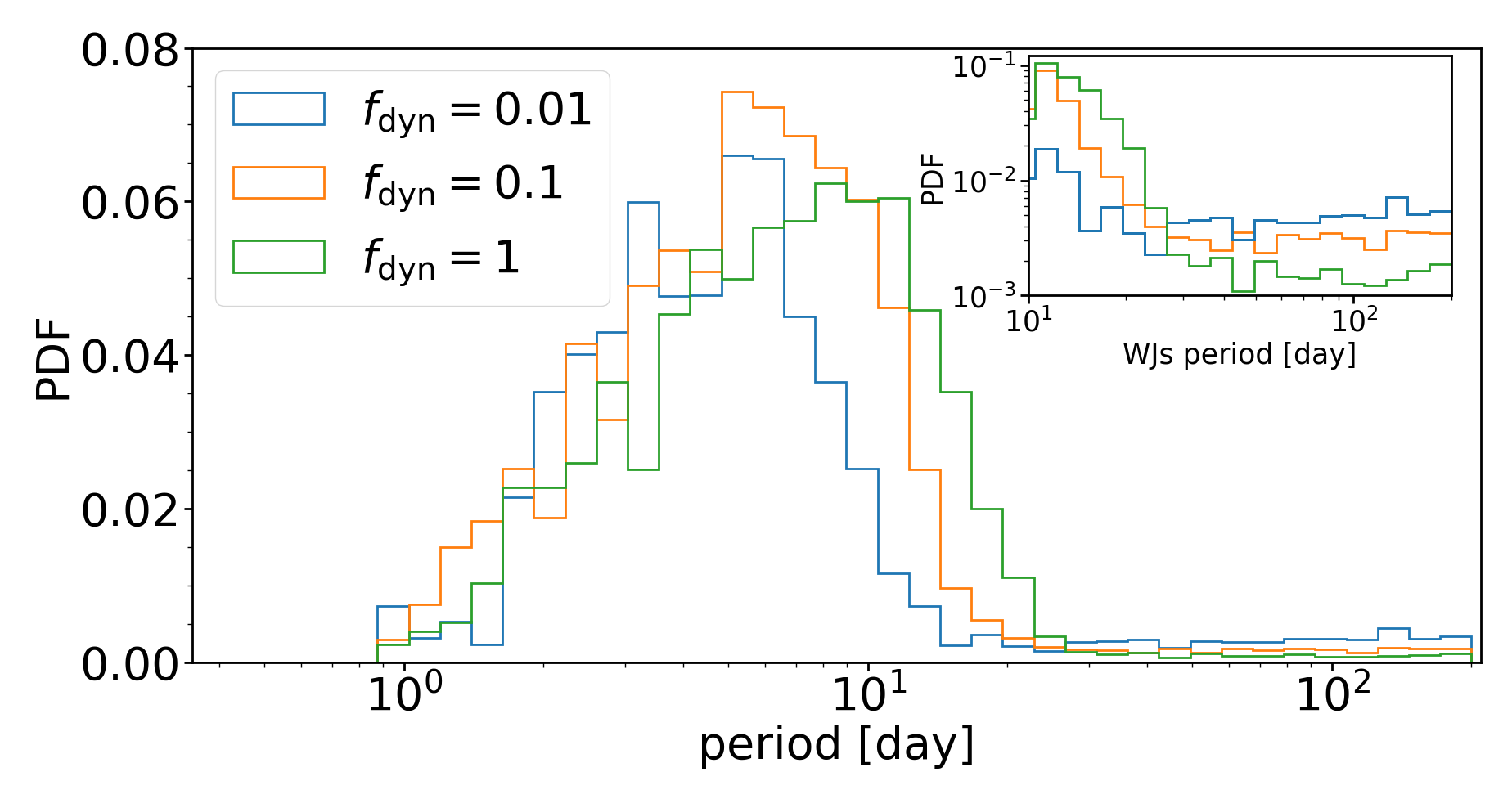}
\caption{ 
The period distribution as derived from the Monte-Carlo simulation, based on the semi-analytical model, for different choices of $f_{\rm dyn}$, normalized to $1$. The rest of the parameters are drawn according $\sigma_1 R_3 a_1 d$, that corresponds to $\sigma_e=0.5$.In the inset figure we introduce the probability distribution function of WJs only (with the same color code)}
\label{fig:period distribution -- fdyn}
\end{figure}

In Fig. \ref{fig:period distribution -- fdyn} we present the dependence on the dynamical tides model on $f_{\rm dyn}$. Larger $f_{\rm dyn}$ corresponds to more efficient extraction of energy via tidal force, which yields to faster tidal migration. In terms of population, the period distribution peak move towards larger periods as the efficiency of the tides rises and there is enhanced tidal disruption, together with the enhanced production of HJs and WJs.  

\begin{figure}
\centering
\includegraphics[width=1\linewidth]{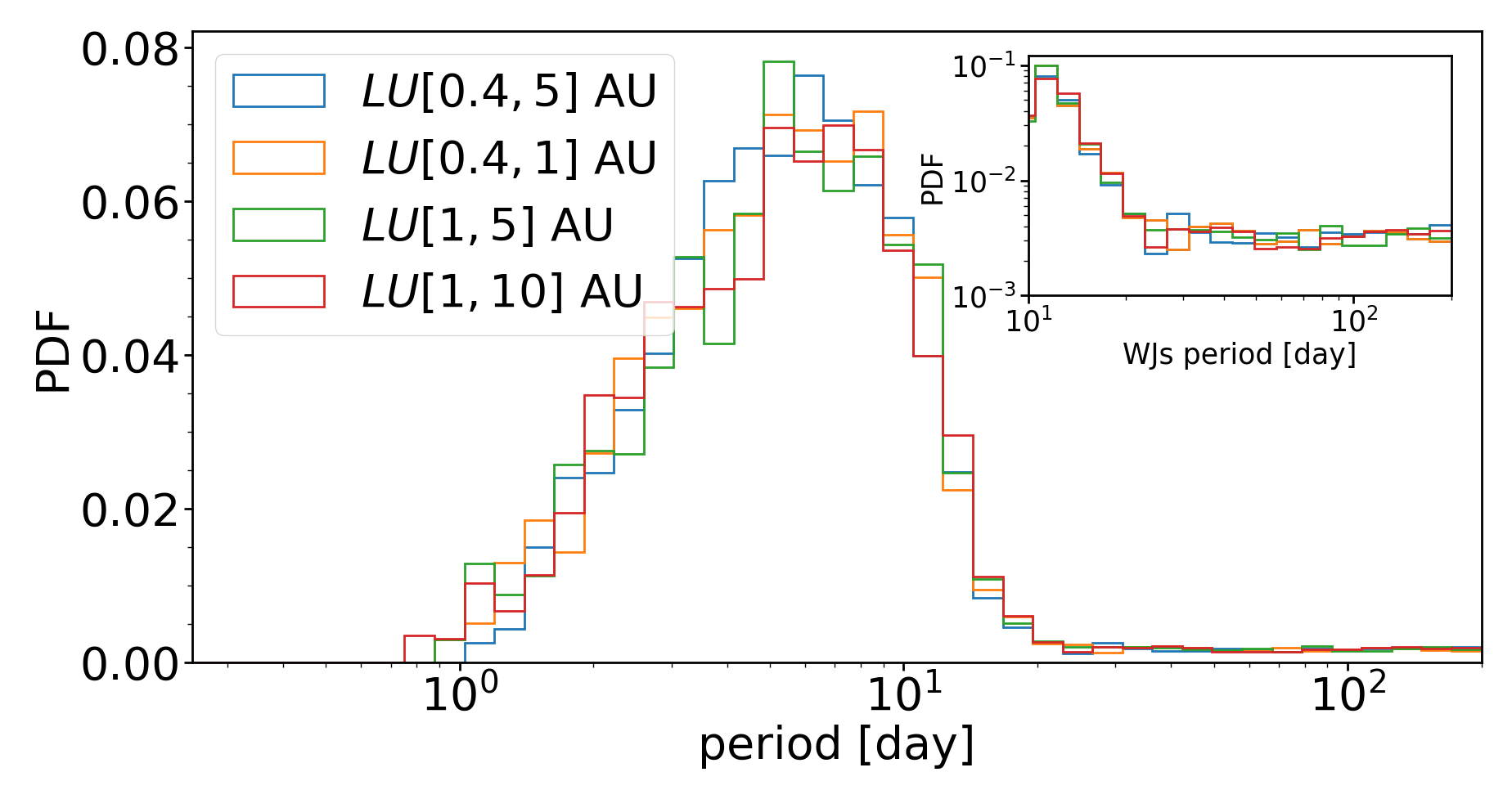}
\caption{ 
The period distribution as derived from the Monte-Carlo simulation, based on the semi-analytical model, for different choices of semimajor axis distribution, normalized to $1$, considering continuous star formation. The rest of the parameters are drawn according $\sigma_1 R_3 df_{0.1}$. $LU$ stands for logarithmic uniform distribution. In the inset figure we introduce the probability distribution function of WJs only (with the same color code)}
\label{fig:period distribution -- sma}
\end{figure}

In Fig. \ref{fig:period distribution -- sma} we present the dependence on our choice of the boundaries in the semimajor axis distribution, and consider the contributions from different semimajor axes. The overall distribution suggests a preference for HJs production from large initial SMAs, however the results seems robust under our choices of distributions.

\section{Discussion \& Implications}\label{sec:discussion & implications}
{\bf Key findings:} As discussed and shown above, inflated eccentric migration following planet-planet scattering in planetary systems, could potentially explain the whole population of eccentric WJs and a significant fraction of HJs or even all, given the lowest inferred estimates).
It also leads to high disruption rate of systems, as it accelerates the parameter space flow from HJs to tidally-disrupted gas-giants.  
Inflated eccentric migration could also play an important role in other systems where high eccentricities are excited through secular processes in general and ZLK oscillations in particular, these will be discussed in future studies (see also \cite{Petrovich2015b} for the study of contracting planets in this context). Our results suggest that any modeling of planetary systems, and in particular young planetary systems, should self-consistently account for the thermal evolution. It should consider the evolving size of the planets and the coupling between the thermal and the dynamical evolution. These could play a key role in the systems dynamics, and in the final sculpting of its architecture. 

The longer overall timescales and shorter time spent at eccentric orbits would potentially allow for more significant contraction of the gas-giants before significant migration occurs, giving rise to a weaker, though still important, effect of the inflated eccentric migration. It should be noted that it is important to model the ZLK-Octupole order, in this context, given the conditions and timescales involved \citep{Naoz2011HJ}.

Inflated eccentric migration is most pronounced at the earliest times when planets are still at their infancy/young-age and are still far more inflated than at later times after contraction. Moreover, when eccentricity excitation occurs through secular processes, tidal effects can induce tidal precession that can quench the eccentricity excitation giving rise to lower eccentricities, that would be expected without the effects of tides, and thereby leading to larger pericenter approaches and less effective tidal dissipation and migration. 

{\bf Implications for giant planet formation:} Our results could have a wide-range of implications in respect to various aspects of giant planet formation and evolution. They shed light and point to the critical role played by the physical evolution of the planets and its coupling to the dynamical evolution, which significantly change the behaviour of eccentric migration processes. It gives rise the efficient formation of both HJs and eccentric WJs, where the latter, in particular, are more difficult to efficiently form through previously studied eccentric migration. We provide their relative fractions overall and as a function of the age of the systems, as well as detailed predictions of their physical properties. Furthermore, the shorter migration timescales due to inflated eccentric migration could give rise to very young, Myrs old HJs, which are typically suggested to form through disk migration (see a detailed review on disk migration in \citealp{Baruteau2014}), or via other channels that focus on the stages after the gas dissipation (e.g. \citealp{Wu2006}).
The HJs\&WJs formed in the proposed channel could have a range of inclinations, even retrograde ones, but would generally have a preference for prograde orbits given their initial inclinations were excited by 
planet-planet scattering \citep{Beauge2012}. 

{\bf Potential caveats and challenges:} In our models, all the WJs formed via eccentric tidal migration are effectively transient HJs that didn't complete their migration within a given time, i.e. eccentric WJs rather than circular ones (although they could reach relatively small eccentricities and even $\lesssim 0.1$), as can be seen in the eccentricities distributions in Figs \ref{fig:Monte Carlo density} and \ref{fig:period distribution -- eccentricity}. Even with the inflated radii, allowing for the formation of lower eccentricity WJs, low ($<0.6$) eccentricity WJs can hardly be formed through inflated migration, and were likely formed in-situ and/or through disk migration, generally consistent with analysis by \cite{AndersonLaiPu2020}. We do note that the paucity of $>0.9$ eccentricity suggested by \cite{Socrates2012a} and ruled out by \cite{DawsonMurrayClay2015} 
does not constrain our models, that indeed show $>0.9$ WJs to be very rare. That being said, the apparent low frequency of $0.6-0.9$ eccentricity WJs is a challenge to inflated eccentric migration, and any other eccentric migration model. In fact, this is a potentially more general difficulty - any successful model for HJ/WJ production should also be able to suppress HJ/WJ formation through the various types of eccentric migration.  

On the theoretical front, both the tidal interaction of gas-giants and the heat transfer to the core are not well understood, giving rise to large uncertainties in the evolutionary models. Here we tried to bracket these potential caveats, but, naturally, better understanding of these processes is critical for the assessment and modeling of any eccentric migration model.   

{\bf Future work on eccentric migration:} Though we focused on the role of inflated migration for specific types of eccentric migration (initialed by planet-planet scattering, and considering some specific models for weak and dynamical tides), the same coupled evolution is important for any suggested eccentric migration model. Follow-up papers may consider other such models and their variants. In addition, accounting for the inflated sizes of gas-giants in their early phases is also important for the increased likelihood for their physical collisions with other planets (due to their larger cross-sections), which in turn can give rise to heating and further inflation of the collision product \citealp{LinIda1997} 
which could then affect migration and further collisions. Eccentric inflated migration could also be coupled to other processes that play a role in planet formation and dynamics, such as photoevaporation (e.g. \citealp{Tripathi2015}). Since the geometric cross-section of initially inflated planets is larger than the cross-section of the non-inflated ones, the role of photoevaporation might change accordingly. 
Furthermore, although we discussed the eccentric tidal migration channel in this paper, initially inflated gas-giants could also be discussed in the context of disk migration, where the radius plays a role in the evolution too.

{\bf Central heating:} We can use our model to set constraints on the effective amount of energy penetrating to the center, since it will affect on the contraction timescale and hence the migration timescale. The exact amount of energy deposits and its distribution, are still unknown and by using our population synthesis, assuming given distribution, e.g. for simplicity, all the energy deposited at the center, we can estimate the energy amount needed to explain the observational results. It should be noted that the parameters space is reach and includes many degeneracies. For example, a choice of some initial eccentricity distribution might shift the distribution in the similar direction in which external heating would do. 

\section{Summary}\label{sec:summary}

In this paper, we proposed the formation channel of hot and warm Jupiters via inflated eccentric migration and discussed the implications on the population. Here, we focused on the semi-analytical approach and discussed specific examples and a detailed population synthesis. We compared the semi-analytic approach with numerical planet evolution models described in paper II,
and find that the semi-analytical model is in good agreement with the numerical model in all the regimes available for the numerical model. This allows us to make use of the more efficient semi-analytic model to explore the dynamics of a large populations of planets and explore the resulting populations of HJs/WJs for a wide range of initial conditions. 

Our models are general and able to include in principle any kind of external heating/dynamical evolution. For brevity, we demonstrated its use for several examples only. 
We presented specific examples for inflated eccentric migration and provided a detailed population synthesis study, based on the semi-analytical model. We studied the dependence of the resulting HJ/WJ populations on the assumptions made regarding the tidal models and heat transfer processes, as well as the initial properties of the progenitor planets and their initial orbits. 

Using the detailed semi-analytic model and population synthesis results we showed that inflated eccentric migration could significantly shorten the migration timescales of HJs and WJs, generally form WJs more efficiently, and give rise to enhanced rates of tidal disruptions of gas-giants. We also considered the effect of external energy injected to the migrating gas-giants on their final formed population and found out that it leads to enhanced tidal disruption, together with smaller fraction of HJs and larger fraction of WJs compared with models without efficient central heating. 

Inflated eccentric migration leads to significant differences in the final distribution of parameters than non-inflated models, and suggest that inflated migration plays an important role in the migration and should generally accounted for in any eccentric migration models (and possibly in disk migration models).

\section*{Acknowledgements}
We gratefully acknowledge fruitful discussions with
Sivan Ginzburg, Thaddeus D. Komacek, Michelle Vick, Nicholas C. Stone and Eden Saig. MR acknowledges the generous support of Azrieli fellowship.




\bibliographystyle{aasjournal}








\end{document}